\documentclass[prb,a4paper,superscriptaddress,twocolumn,showpacs,final]{revtex4}

\usepackage{graphicx}
\usepackage{fancyhdr}
\usepackage{afterpage}
\usepackage{fancyvrb}
\usepackage[amssymb]{SIunits}
\usepackage[centertags]{amsmath}
\usepackage{nomencl}
\usepackage{epsfig}
\usepackage{amsfonts}
\usepackage{amsthm}
\usepackage{latexsym}
\usepackage{dcolumn}
\usepackage{epic}
\usepackage{hyperref}
\usepackage{color}
\usepackage{newlfont}
\usepackage{mathrsfs}
\usepackage{ulem}

\DeclareGraphicsExtensions{.ps,.eps,.pdf,.jpeg,.jpg}

\newcommand{\nn}{\nonumber}

\newcommand{\Eq}[1]{Eq.~(\ref{#1})}

\newcommand{\dg}{^\dagger}
\newcommand{\pdg}{^{\phantom{\dagger}}}

\newcommand{\smallfrac}[2]{\mbox{$\frac{#1}{#2}$}}
\newcommand{\half}{\smallfrac{1}{2}}
\newcommand{\bra}[1]{\langle{#1}|}
\newcommand{\ket}[1]{|{#1}\rangle}
\newcommand{\sq}[1]{\left[ {#1} \right]}
\newcommand{\cu}[1]{\left\{{#1} \right\}}
\newcommand{\ro}[1]{\left( {#1}\right)}
\newcommand{\an}[1]{\left\langle{#1}\right\rangle}
\newcommand{\st}[1]{\left|{#1}\right|}
\newcommand{\tr}[1]{\mathrm{Tr}\sq{ {#1}}}

\newcommand{\du}{\partial}

    \newcommand{\V }{{\cal V}}
    \newcommand{\W }{{\cal W}}
    \newcommand{\e }{{e}}
    
    \newcommand{\D}[1]{{\cal D}\sq{{#1}}}

    \newcommand{\h}[1]{{\cal H}\sq{{#1}}}
    
    \newcommand{\hlin}[1]{\bar{\cal H}\sq{{#1}}}

    \newcommand{\Ref}[1]{Ref.~\onlinecite{#1}}
    \newcommand{\Refs}[1]{Refs.~\onlinecite{#1}}
    \newcommand{\Fig}[1]{Fig.~\ref{#1}}
    \newcommand{\Sec}[1]{Sec.~\ref{#1}}
    \newcommand{\sig}{\hat{\sigma}}
    
    \newcommand{\etal}{\textit{et al.}}

    \newcommand{\rmd}{\mathrm{d}}
    \newcommand{\rhoc}{\rho_{\mathrm{c}}}
    \newcommand{\rhocbar}{\bar{\rho}_{\mathrm{c}}}

\newcommand{\rf}{\mathrm{rf}}

\newcommand{\TL}{\mathrm{TL}}
\newcommand{\rL}{\mathrm{L}}

\newcommand{\dt}{{\rmd t}}
\newcommand{\Ham}{\hat{H}}

\newcommand{\ortega}{\alpha}
\newcommand{\Tcal}{\mathcal{T}}
\newcommand{\Xcal}{\mathcal{X}}
\newcommand{\nlalign}[1]{\nn\\&\phantom{=}{#1}}

\newcommand{\var}{\mathrm{v}}

\begin{document}
\title{Model for monitoring of a charge qubit using a 
radio-frequency quantum point contact including experimental 
imperfections}

\author{Neil P. Oxtoby}
\email{n.oxtoby@herts.ac.uk}
\affiliation{%
Centre for Quantum Computer Technology,
Centre for Quantum Dynamics,
School of Biomolecular and Physical Sciences,
Griffith University,
Brisbane 4111,
Australia
}%
\affiliation{%
Quantum Physics Group, STRI, 
School of Physics, Astronomy and Mathematics, 
University of Hertfordshire, Hatfield, Herts AL10 9AB, 
United Kingdom
}%

\author{Jay Gambetta}
\affiliation{Institute for Quantum Computing 
and Department of Physics and Astronomy, 
University of Waterloo, Waterloo, 
Ontario N2L 3G1, Canada}%
\affiliation{%
Department of Applied Physics and 
Department of Physics,
Yale University, New Haven, Connecticut 06520, USA}%

\author{H. M. Wiseman}
\email{h.wiseman@griffith.edu.au}
\affiliation{%
Centre for Quantum Computer Technology,
Centre for Quantum Dynamics,
School of Biomolecular and Physical Sciences,
Griffith University,
Brisbane 4111,
Australia
}%

\begin{abstract}
{The extension of quantum trajectory theory to 
incorporate realistic imperfections in the measurement of 
solid-state qubits is important for quantum computation, 
particularly for the purposes of state preparation and 
error-correction as well as for readout of computations.  
Previously this has been achieved for low-frequency 
(dc) weak measurements.  In this paper we extend realistic 
quantum trajectory theory to include radio frequency (rf) 
weak measurements where a low-transparency quantum point 
contact (QPC), coupled to a charge qubit, is used to damp a 
classical oscillator circuit.  }  
The resulting realistic quantum trajectory equation 
must be solved numerically.  We present an analytical 
result for the limit of large dissipation within the 
oscillator (relative to the QPC), where the oscillator 
slaves to the qubit.  
The rf+dc mode of operation is considered.  
Here the QPC is biased (dc) as well as subjected to a 
small-amplitude sinusoidal carrier signal (rf).  
The rf+dc QPC is shown to be a low-efficiency charge qubit 
detector, that may nevertheless be higher than the dc-QPC 
(which is subject to $1/f$ noise).
\end{abstract}

\date{published 5 March 2008: 
\href{http://dx.doi.org/10.1103/PhysRevB.77.125304}{\underline{link}}}

%\doi{10.1103/PhysRevB.77.125304}

\maketitle

%********************************************************
\section{Introduction\label{sec:intro}}
{Solid-state proposals for building scalable quantum information 
processors\cite{KanNAT98,LosDiVPRA98,PriVagKvePLA98,ImaetalPRL99,VrietalPRA00} 
seem promising.  
In any quantum information processor, the quantum bits (qubits) 
of information need to be read out, as well as controlled (via 
measurement-based feedback, for example).  
Quantum trajectory theory\cite{OpenSys,WisMilPRA93b,WisQSO96} 
has been used to describe single realizations of the 
continuous-in-time weak quantum measurement of electronic charge 
qubits\cite{KorPRB99,KorPRB01b,WisetalPRB01,GoaMilWisSunPRB01,
KorPRB01c,GoaMilPRB01,KorPRB03,Kor03,StaBar-04,StaBarPRL04,
GoaPRB04} \textit{conditioned} by the electrical output of a 
mesoscopic measurement device such as a quantum point contact 
(QPC) or single-electron transistor (SET).  
{Very recently, quantum trajectory theory has been 
applied to circuit QED.\cite{GametalPRA08} }  
In these works the qubit evolution was conditioned on 
idealized measurement results (such as electron tunnelling) 
at the scale of the mesoscopic detector.  That is, the 
observer's state of knowledge about the qubit state is 
updated based on these idealized measurement results.  
Two of us recently extended such work to condition the 
qubit state on a macroscopic signal that is realistically 
available to an observer in a dc-QPC measurement.\cite{OxtetalPRB05}  
In particular, this extension considered the noisy, filtering 
characteristic of an external circuit, including an amplifier.  
The result is a corrupted version of the idealized measurement 
results upon which the qubit evolution can be conditioned.  
This extension is known as ``realistic quantum trajectory'' 
theory and was pioneered for photodetection in quantum 
optics,\cite{WarWisMabPRA02,WarWisJOBQSO03a} 
where it was applied to an avalanche photodiode and a 
photoreceiver.
}

In traditional dc charge-qubit measurement techniques, low 
frequency noise (``$1/f$ noise''\cite{ZimetalAPL92,WonMR03}) 
limits the detector sensitivity.\cite{SchetalSC98}  
To circumvent this, Schoelkopf 
\etal\cite{SchetalSC98} introduced the so-called 
radio-frequency single-electron transistor (rf-SET).  
For the original configuration, the rf-SET demonstrated 
constant gain from dc to 100MHz --- an improvement on 
the conventional SET bandwidth by two orders of 
magnitude.\cite{SchetalSC98}  
The idea is to measure the damping of a resonant 
(oscillator) circuit in which the SET is embedded.  
In the context of charge-qubit detection,\cite{AasetalPRL01} 
the damping depends on the qubit state, via the SET.  
Thus, monitoring the damping of the oscillator constitutes 
a continuous-in-time measurement of the charge qubit.  
This concept can be applied to any charge-sensitive 
detector, in particular to the QPC, for 
example.\cite{QinWilAPL06} 

%Some previous work\cite{TurKorAPL03,TurKorPRB04} has been 
%done on calculating the charge sensitivity and response of 
%the rf-SET, as well as various other work that is not of 
%particular relevance to this paper.  
%However, to our knowledge, there are no continuous 
%measurement theories describing the conditional evolution 
%of a charge qubit state monitored by a rf-SET, nor a rf-QPC.  

In this paper we derive an evolution equation for the 
conditional state of a charge qubit monitored continuously in 
time by a detector operated in the rf configuration. To the 
best of our knowledge, an equation of this type has not 
previously been derived.  
We also consider conditioning the qubit state on 
measurement results available to a realistic observer, 
within the framework of realistic quantum trajectory 
theory.\cite{WarWisMabPRA02,WarWisJOBQSO03a,OxtetalPRB05,
GamWisJOBQSO05}  
In this approach, the bare charge-qubit detector 
(QPC in our case) is embedded in a realistic circuit, and 
an equation is derived that describes the evolution 
of the combined circuit-plus-qubit state conditioned on 
measurement results available to a realistic observer.

{
Being able to determine the state of a quantum system 
conditioned on actual measurement results is 
expected to be vitally important for quantum computation, 
particularly for state preparation and quantum error 
correction.\cite
{AhnWisMilPRA03,SarAhnJacMilPRA04,AhnWisJacPRA04,vHanMab05}  
More broadly, it is also essential for understanding and 
designing optimal feedback control.\cite
{WisMilPRL93,WisPRL95,DohJacPRA99,DohetalPRA00,
ArmetalPRL02,WisManWanPRA02,KorPRB01b,SmietalPRL02,
RusKorPRB02,JacPRA03,ComJacPRL06}
}

To simplify our analysis, we make a number of approximations.  
First, we use the low-transparency QPC as the charge-sensitive 
detector embedded in the rf circuit, instead of the SET.  
Second, we make a rotating wave approximation (RWA) 
to remove the $1/f$ noise from our equations (as the 
rf configuration removes it in the experiment).  
In order to do this, we assume the QPC to be operating 
in the weakly-responding (diffusive) 
limit.\cite{KorPRB99,KorPRB01b,GoaMilWisSunPRB01}  
{ In this limit, the QPC shot noise appears as white 
noise, which is modeled as a Gaussian-distributed 
stochastic process (having a diffusive appearance).  }  
Third, we assume that the rf-QPC is operated in the rf+dc 
mode introduced in the context of a SET in 
\Ref{AasetalAPL01}.  This is where the QPC is subjected 
to a small amplitude sinusoidal oscillation (rf) superposed 
on a relatively large bias (dc).  
We find that the rf+dc QPC is a highly inefficient 
charge-qubit detector.  However, this may be higher than 
the measurement efficiency of the dc-QPC, which in practice 
is not only limited by $1/f$ noise, but is also further 
degraded the longer the measurement has to continue.

The paper is organized as follows.  
The qubit and QPC are discussed in \Sec{sec:quantum}.  
The classical circuit is discussed and analyzed 
in \Sec{sec:classical}, including the presentation of 
stochastic differential equations describing the state 
of the oscillator circuit.  
The stochastic master equation for the qubit state conditioned on 
the bare detector output is presented in \Sec{sec:qubit}.  
The realistic quantum trajectory derivation then proceeds 
in an analogous manner to \Ref{OxtetalPRB05}, where the 
circuit state was described by one parameter --- the 
charge on a capacitor.  In general the oscillator circuit needs two 
parameters to describe its state, thus hinting at a 
complication in deriving the realistic equations for 
the rf configuration.  
{ 
However, because the QPC only damps the oscillator circuit 
(and doesn't induce a phase-shift in the reflected signal), 
all of the qubit information is contained in the amplitude 
(damping) of the reflected signal.  Therefore we can proceed 
with the realistic quantum trajectory derivation using a 
single parameter to describe the state of the classical circuit.}  
As always, numerical calculations must 
be performed to obtain the realistic quantum trajectories.  
However, in \Sec{sec:slaved} we present analytical results 
for the heavily damped limit where the circuit is 
slaved to the qubit (adiabatic elimination of the 
circuit).  
The paper is concluded in \Sec{sec:conclusion}.

%********************************************************
\section{Qubit and QPC\label{sec:quantum}}
\begin{figure}[!ht]
\begin{center}
\includegraphics[width=0.4\textwidth]{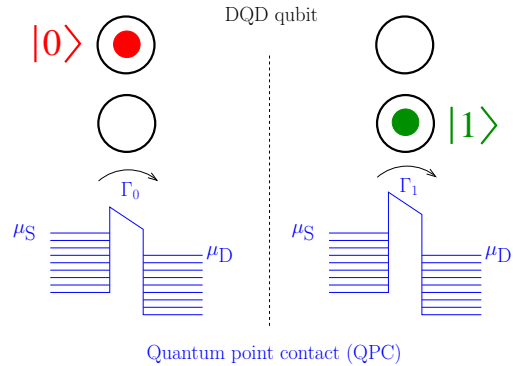}
\end{center}
\caption[Schematic of an isolated DQD qubit and QPC]
{\label{fig:dqdqpc}
Schematic of an isolated DQD qubit and capacitively coupled 
low-transparency QPC between source (S) and drain (D) leads.  
}
\end{figure}

The measured quantum system we consider is the 
double-quantum-dot (DQD) charge 
qubit.\cite{HayetalPRL03,GorHasWilPRL05}  
A schematic of the isolated DQD \cite{GorHasWilPRL05} and 
capacitively coupled QPC is shown in \Fig{fig:dqdqpc}.  
We consider the \textit{low-transparency} QPC, and represent it as 
a tunnel barrier between source and drain leads (reservoirs) 
with respective Fermi levels $\mu_\mathrm{S}$ and 
$\mu_\mathrm{D}$.  The QPC voltage bias is 
$\e V_{\rmd}=\mu_\mathrm{S}-\mu_\mathrm{D}$, where 
$\e > 0$ is the quantum of electronic charge.%
\footnote{The choice $\e>0$ corresponds to defining current 
          in terms of the direction of electron flow.  
          That is, in the opposite direction to conventional 
          current.}  
The DQDs are occupied by a single excess electron, the 
location of which determines the charge state of the qubit.  
The charge basis states are denoted $\ket{0}$ and $\ket{1}$ 
(see \Fig{fig:dqdqpc}).  
We assume that each quantum dot has only one single-electron 
energy level available for occupation by the qubit electron, 
denoted by $E_{1}$ and $E_{0}$ for the near and far dot, 
respectively.  

%%%%%%%%%%%%%%%%%%%%%%%%%%%%%%%%%
%%%%%%%%%%%%%%%%%%%%%%%%%%%%%%%%%
%%%%%%%%%%%%%%%%%%%%%%%%%%%%%%%%%
%%%%%%%%%%%%%%%%%%%%%%%%%%%%%%%%%

The Hamiltonian for the qubit can be written as 
\begin{equation}
\label{eq:Hqb}
\Ham_\mathrm{qb} = 
\frac{1}{2}\ro{\varepsilon\sig_{z} + \Omega_\mathrm{tun}\sig_{x}} ,
\end{equation}
where the qubit energy asymmetry is 
$\varepsilon \equiv E_{1}-E_{0}$, $\Omega_\mathrm{tun}$ is the 
DQD tunnel-coupling strength, 
and $\sig_{x,z}$ are Pauli matrices in the measurement 
(charge) basis.  
The eigenvalues of $\Ham_\mathrm{qb}$ are $\mp \Omega/2$, 
where $\Omega \equiv \sqrt{\Omega_\mathrm{tun}^2 + \varepsilon^2}$.  

Associated with each of the qubit charge states is a current 
through the detector.  The average current through the 
detector is $I_{1} = \e\Gamma_{1}$ when the nearby dot is 
occupied, and $I_{0} = \e\Gamma_{0}$ when the nearby dot is 
unoccupied.  
The variation in the detector output that depends on the qubit 
state is referred to as the detector's {\textit{response}} 
and is denoted $\Delta I \equiv I_{1}-I_{0}$.  
We can quantify the strength of the detector response by 
$0 \leq \st{\Delta I}/{I_\mathrm{av}} \leq 2 $, 
where ${I_\mathrm{av}} \equiv \ro{I_{1}+I_{0}}/2$.  Thus, a 
weakly responding\cite{KorPRB99} %,KorPRB01b,Kor03} 
detector satisfies 
$\st{\Delta I} \ll {I_\mathrm{av}}$, and a detector with finite, or 
strong, response satisfies $\st{\Delta I} \sim {I_\mathrm{av}}$.  
{In this paper we consider the limit of a weakly responding QPC, 
where the QPC shot noise appears as a diffusive, white noise 
process.}

The total Hamiltonian for the system is
\begin{equation}
\Ham_\mathrm{Tot} = \Ham_\mathrm{qb} + \Ham_\mathrm{F} 
               + \Ham_\mathrm{T} + \Ham_\mathrm{coup} ,
\label{eq:Htotal:qpc}
\end{equation}
where the qubit Hamiltonian $\Ham_\mathrm{qb}$ is given by \Eq{eq:Hqb}.  
The free Hamiltonian describing the continua of electron channels 
(momenta) $k$ and $q$ in the source and drain leads is 
\begin{equation}
\Ham_\mathrm{F} = 
 \sum_{k}\omega\pdg_{\mathrm{S}k}
         \hat{a}\dg_{\mathrm{S}k}\hat{a}\pdg_{\mathrm{S}k}
+\sum_{q}\omega\pdg_{\mathrm{D}q}
         \hat{a}\dg_{\mathrm{D}q}\hat{a}\pdg_{\mathrm{D}q} ,
\label{eq:HF:qpc}
\end{equation}
where $\hat{a}_{\mathrm{S}}$ and $\hat{a}_{\mathrm{D}}$ are the 
Fermi field annihilation operators for the source and drain 
leads, respectively.  
The tunnelling Hamiltonian 
\begin{equation}
\Ham_\mathrm{T} = 
\sum_{k,q} T^{\phantom{*}}_{kq}
           \hat{a}\dg_{\mathrm{S}k}\hat{a}\pdg_{\mathrm{D}q}
+T^{*}_{qk}\hat{a}\dg_{\mathrm{D}q}\hat{a}\pdg_{\mathrm{S}k}
\label{eq:HT:qpc}
\end{equation}
describes tunnelling between the source and drain leads.  
The probability amplitude for a source electron in channel $k$ 
to tunnel through the QPC into the drain channel $q$ is 
$T^{\phantom{*}}_{kq}$.  The coupling Hamiltonian 
\begin{eqnarray}
\Ham_\mathrm{coup} = \hat{n}
\ro{\sum_{k,q}\chi^{\phantom{*}}_{kq}
    \hat{a}\dg_{\mathrm{S}k}\hat{a}\pdg_{\mathrm{D}q}
   +\chi^{*}_{qk}
    \hat{a}\dg_{\mathrm{D}q}\hat{a}\pdg_{\mathrm{S}k}
    }
\label{eq:Hcoup:qpc}
\end{eqnarray}
describes the change in the effective QPC tunnelling amplitude 
from $T_{kq}\rightarrow T_{kq}+\chi_{kq}$ when the nearby dot is 
occupied.  This changes the QPC current from 
$I_{0} = e\st{\cal T}^{2}$ to 
$I_{1} = e\st{{\cal T} + {\cal X}}^{2}$.  Here 
${\cal T}\propto T_{kq}$ and ${\cal X}\propto \chi_{kq}$ 
are both proportional to the square-root of the source-drain 
voltage $V_{\rmd}$.\cite{GoaMilWisSunPRB01}%
$^,${\footnote{
The work of B\"uttiker in \Ref{ButPRB92}, for example, 
is a more general reference for QPC conductance calculated using 
scattering theory.}}
The occupation number operator of the nearby dot is 
$\hat{n}=(1+\sig_{z})/2$.  
Note that the height of the QPC (tunnel-junction) barrier 
is increased when the nearby dot is occupied, 
due to electrostatic repulsion, so that $I_{0} > I_{1}$.

%********************************************************
\section{Classical System: Oscillator\label{sec:classical}}
\begin{figure}[ht]
\begin{center}
\includegraphics[width=0.48\textwidth]{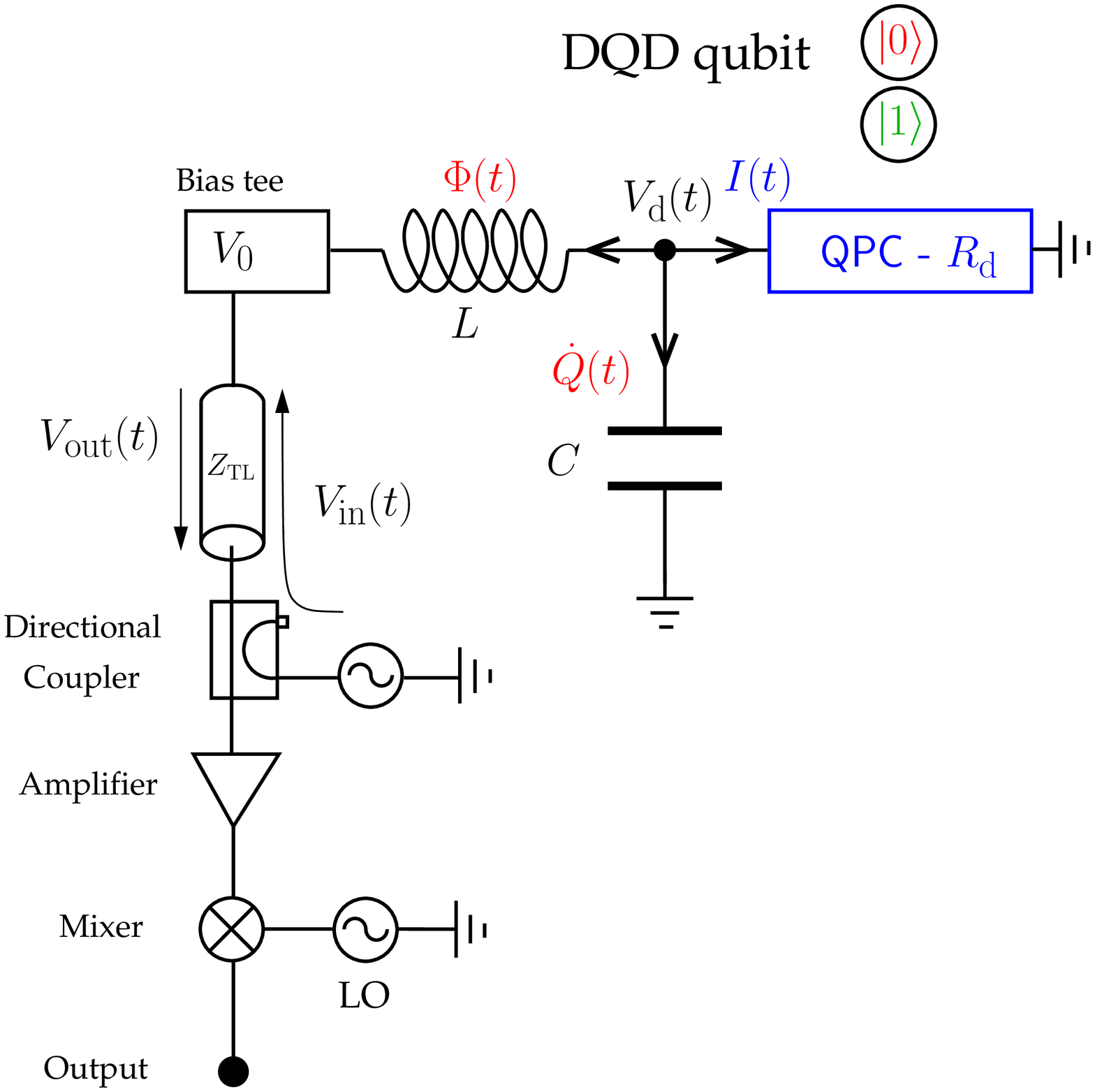}
\caption[Equivalent circuit for rf detection]{%
Equivalent circuit for continuous monitoring of a charge qubit 
coupled to a classical $LC$ oscillator with inductance $L$ and 
capacitance $C$.  
We consider the charge-sensitive detector that loads the 
oscillator circuit to be a QPC (see Fig. \ref{fig:dqdqpc} 
for details).  Measurement is achieved using reflection 
with the input voltage, $V_\mathrm{in}(t)$, and the output 
voltage, $V_\mathrm{out}(t)$, being separated by a directional 
coupler.  The output voltage is then amplified and mixed with 
a local oscillator, $LO$, and then measured.  
\label{fig:rfcircuit}}
\end{center}
\end{figure}

An oscillator circuit, or tank circuit, consists 
of an inductor $L$, and capacitor $C$.  
We treat the oscillator classically.  
The (angular) frequency for which resonance occurs in 
such an unloaded tank circuit, $\omega_{0}=1/\sqrt{LC}$, 
is known as the {\textit resonance} frequency.  
Embedding a dissipative component, like a resistor, into 
the tank circuit provides a source of damping.  
For our purposes, the QPC provides the damping, 
so the oscillator damping depends on the qubit state 
{via the qubit-dependent QPC conductance}.  
It therefore makes sense to monitor the damping of the tank 
circuit in order to ascertain the qubit state.  
This is achieved by using the tank circuit to terminate a 
transmission line.  Impedance mismatch between the tank 
circuit and transmission line causes a signal launched 
towards the tank circuit to be reflected back along 
the transmission line, where the reflected signal can be 
observed.\footnote{There is also a technique involving the 
   transmission of rf radiation.  See Refs. 
   \onlinecite{QinWilAPL06}, \onlinecite{FujHirAPL00}, and 
   \onlinecite{CheetalAPL02}, for example.}  
See \Fig{fig:rfcircuit} for our equivalent circuit 
representing the rf-QPC setup.  
A recent experiment\cite{QinWilAPL06} reports realization 
of a semiconductor rf-PC (radio-frequency point contact), 
with two benefits over the SET --- lower PC impedance 
(thus simplifying impedance matching with the transmission 
line), and easier fabrication.  This as yet unoptimized 
device exhibits a lower charge sensitivity than the rf-SET.  
In said experiment, the point contact is operating as a 
simple voltage-controlled resistor rather than a QPC.  
{Another more recent experiment\cite{ReietalAPL07} has 
realized fast charge sensing with a semiconductor rf-QPC.}

The AC voltage signal launched towards the tank circuit 
is referred to as the ``carrier''.\cite{SchetalSC98}  
Using a carrier signal frequency equal to the resonance 
frequency of the unloaded tank circuit allows the detector 
to be replaced by its (frequency-dependent) dynamic resistance 
$R_{\rmd}$.\cite{AasetalAPL01,TurKorPRB04} 
This is a valid first-order approximation since the tank 
circuit is most sensitive to frequencies within a small 
bandwidth around the resonance frequency.  

Consider the equivalent circuit of \Fig{fig:rfcircuit}.  
The oscillator circuit consisting of an inductance $L$ and 
capacitance $C$ terminates the transmission line of impedance 
$Z_{\TL}=50\Omega$.  The voltages (potential drops) across 
the oscillator components can be written as 
\begin{subequations}\begin{align}
V_{\rL}(t) &= \dot{\Phi}(t) ,\label{eq:VL} \\
V_\mathrm{C}(t) &= \frac{Q(t)}{C} ,\label{eq:VC}
\end{align}\label{eqs:VLVC}\end{subequations}
where $\Phi(t)$ is the flux through the inductor, and $Q(t)$ 
is the charge on the capacitor.  These variables represent 
conjugate variables that together completely characterize the 
classical oscillator state.  
The voltage across the detector, $V_{\rmd}(t)$, is the topic 
of the following subsection.  The current flowing through the 
detector, $I(t)$, will be discussed in \Sec{sec:qubit}.  
We define the incoming (relative to the tank circuit) 
transmission line voltage as the carrier signal 
$V_\mathrm{c}(t)=V^{\rf}_\mathrm{in}\cos(\omega_{0}t)$
plus some noise $N(t)$:
\begin{equation}
V_\mathrm{in}(t) = V_\mathrm{c}(t) + N(t) .
\label{eq:Vin}
\end{equation}
The outgoing voltage in the transmission line is denoted 
$V_\mathrm{out}(t)$.  
As in the experiments of \Ref{BlaetalNJP05}, the detector 
bias $V_{0}$ is added at a bias tee at the end of the 
transmission line.

%%%%%%%%%%%%%%%%%%%%%%%%%%%%%%%%%%%%%%%%%%%%%%%%
%%%%%%%%%%%%%%%%%%%%%%%%%%%%%%%%%%%%%%%%%%%%%%%%
%%%%%%%%%%%%%%%%%%%%%%%%%%%%%%%%%%%%%%%%%%%%%%%%
%%%%%%%%%%%%%%%%%%%%%%%%%%%%%%%%%%%%%%%%%%%%%%%%

%*************************************************
\subsection{Detector voltage 
\label{sec:Vd}}
In \Sec{sec:qubit}, we will discuss the conditional 
dynamics of the quantum system undergoing rf-QPC monitoring.  
Since the QPC tunnelling rates depend on the detector 
voltage $V_{\rmd}(t)$ (see \Ref{GoaMilWisSunPRB01}), 
we calculate it here.  

Using complex phasor notation, the voltage divider rule 
gives the detector voltage as
\begin{equation}
V_{\rmd}(t) = 
  V_{0}
+ \mathrm{Re}\sq{\frac{Z_\mathrm{RC}}{Z}V(t)} ,
\label{eq:divider1}
\end{equation}
since the inductor is a ``short-circuit'' at dc.  
The impedances here are 
$Z_\mathrm{RC}^{-1} = R_{\rmd}^{-1}+i\omega_{0}C$ 
($R_{\rmd}C$ combination), and 
$Z = Z_\mathrm{RC} + Z_\mathrm{L}$ (entire tank circuit), 
with $Z_\mathrm{L}=i\omega_{0}L$ (inductor).  
Here the AC voltage in the transmission line, 
\begin{equation}
V(t) = \ro{1+\beta}V_\mathrm{c}(t) ,
\label{eq:reflected}
\end{equation}
consists of the incident and reflected AC signal.  
Here we make an assumption that $N(t)$ 
comprises predominantly low-frequency noise which will be 
removed by the rf-QPC, and so we drop it to simplify the 
analysis.  
The reflection coefficient, $\beta$, is given in 
terms of the impedance mismatch between tank circuit 
and transmission line as 
\begin{equation}
\beta = \frac{Z-Z_{\TL}}{Z+Z_{\TL}} .
\label{eq:beta}
\end{equation}
That is, the AC signal reflected off the loaded tank 
circuit is given by $\beta V_\mathrm{c}(t)$.  Combining 
these results, the detector voltage can be written as 
\begin{align}
V_{\rmd}(t) &= 
  V_{0}
+ 2\mathrm{Re}\sq{\frac{Z_\mathrm{RC}}{Z+Z_{\TL}}V_\mathrm{c}(t)} 
\nn \\ 
&= 
V_{0}
+ 2\mathrm{Re}\sq{G e^{i\phi}V_\mathrm{c}(t)} .
\label{eq:divider2.2} 
\end{align}
{Here the amplitude gain $G$ and phase-shift $\phi$ of the 
carrier signal when it reaches the QPC are given in terms 
of the circuit and detector quality factors as
\begin{subequations}\begin{align}
G & \equiv
\ro{Q_{\rmd}^{-2}Q^{-2} + Q_\mathrm{T}^{-2}}^{-1/2} ,
\label{eq:G} 
\\
\tan\ro{\phi} &= Q_\mathrm{T} .
\label{eq:phi}
\end{align}\label{eqs:Dphi}\end{subequations}
The quality factors of the unloaded tank circuit, detector, and 
loaded tank circuit are
$Q \equiv \omega_{0}/\gamma$, %~ = Z_\mathrm{LC}/Z_\mathrm{TL} 
$Q_{\rmd} \equiv \gamma_{\rmd}/\omega_{0}$, %~ = R_{\rmd}/Z_\mathrm{LC}
and $Q_\mathrm{T}^{-1} \equiv Q^{-1} + Q_{\rmd}^{-1}$, 
respectively.
}

{In the high quality limit $Q_\rmd, Q \gg 1$ 
($\gamma_\rmd \gg \omega_0 \gg \gamma$), 
the tank circuit damping is due primarily to the 
detector.  This limit therefore represents the highest 
sensitivity for the rf+dc QPC.  We find that 
%$Q^{-2}Q_{\rmd}^{-2}\ll Q_\mathrm{T}^{-2}$, 
$G \sim Q_\mathrm{T}$ % \sim Q/2$ 
and $\phi \sim \pi/2$, so that the voltage across the QPC 
experiences a negative $\pi/2$ phase-shift ($\cos \to \sin$) 
after the inductor.  This is expected since the voltage 
across an (ideal) inductor {\textit lags} the driving signal 
by $\pi/2$.  In this limit the voltage across the QPC is 
}\begin{equation}
V_{\rmd}(t) = 
 V_{0}
 \sq{1 + \epsilon_\mathrm{in}\sin\ro{\omega_{0}t}} .
\label{eq:Vd}
\end{equation}
Here we have defined the dimensionless parameter 
$\epsilon_\mathrm{in} \equiv 
2Q_\mathrm{T}V_\mathrm{in}^{\rf}/V_{0}$, which satisfies 
$\epsilon_\mathrm{in} \ll 1$ in the rf+dc mode.\cite{AasetalAPL01}  
Equation (\ref{eq:Vd}) is in agreement with \Ref{TurKorPRB04}.

%************************************************************
\subsection{Idealized classical dynamics
\label{sec:oscillator}}
The two conjugate parameters we use to describe the 
oscillator state are the flux through the inductor, 
$\Phi(t)$, and the charge on the capacitor, $Q(t)$.  
The dynamics of the oscillator are found by analyzing the 
equivalent circuit of \Fig{fig:rfcircuit} using the 
well-known Kirchhoff circuit laws.  
Doing this we find that the classical system obeys the 
following set of coupled differential equations
\begin{subequations}\begin{align}
\dot \Phi(t)& = -\gamma\Phi(t) + \frac{Q(t)}{C} - V_0 - 2 \sq{V_{\mathrm{c}}(t)+ N(t)},
\label{eq:phidot}\\
\dot Q(t)&=-\frac{\Phi(t)}{L} - I(t),
\label{eq:Qdot}\\
V_{\mathrm{out}}(t) &= V_{\mathrm{c}}(t) + N(t) 
+ \frac{Z_{\TL}\Phi(t)}{L}, \label{eq:vout}
\end{align}\label{eqs:SDEs}\end{subequations} 
where $I(t)$ is detector current and we have re-included the 
input noise $N(t)$.  
{These are the equations of a damped harmonic oscillator 
driven by both the input voltage and the QPC current (including 
the shot noise) at frequency $\omega_0$.  The $1/f$ noise 
in the input signal and the QPC is filtered out.}  
To see this we start by recasting the problem in terms of the 
following dimensionless parameters
\begin{subequations}\begin{align}
 x(t) &\equiv 
%\sqrt{\frac{1}{\hbar L\omega_{0}}} 
\sqrt{\frac{1}{\hbar Z_\mathrm{LC}}}
\sq{\Phi(t)-\Phi_{\mathrm{ss}}}
\label{eq:x} 
 \equiv 
 \ortega_\mathrm{x} \sq{\Phi(t)-\Phi_{\mathrm{ss}}} ,
 \\
 y(t) &\equiv 
%\sqrt{\frac{L\omega_{0}}{\hbar}} 
\sqrt{\frac{Z_\mathrm{LC}}{\hbar}}
\sq{Q(t)-Q_{\mathrm{ss}}} 
 \label{eq:y} 
 \equiv 
 \ortega_{\mathrm{y}} \sq{Q(t)-Q_{\mathrm{ss}}} , 
 \end{align}\label{eqs:xy}\end{subequations}
where $\alpha_\mathrm{x}$ and $\alpha_\mathrm{y}$ are 
implicitly defined above and the subscript ss refers to 
the steady state.  The resulting equations are 
\begin{subequations}\begin{align}
 \dot{x}(t) &= 
  - \gamma x(t) 
  + \omega_{0}y(t) 
  - 2\ortega_\mathrm{x} [V_\mathrm{c}(t)+N(t)],
  \label{eq:xdot}
  \\
  \dot{y}(t) &= 
  - \omega_{0}x(t) 
  - \ortega_{\mathrm{y}}\tilde{I}(t) , 
  \label{eq:ydot}
  \end{align}\label{eqs:xydot}\end{subequations} 
where $\tilde{I}(t) = I(t)- I_\mathrm{ss}$.  
{The solution of these coupled equations involves 
the two time-scales $\omega_0^{-1}$ and $\gamma^{-1}$.  
In the limit $\gamma^{-1} \gg \omega_0^{-1} $ ($Q\gg 1$),} 
we can define a coarse-graining\cite{WisMilPRA93a} 
time-scale $\rmd t$ (Roman font d) that is short compared 
to $\gamma^{-1}$, and long compared to $\omega_{0}^{-1}$.  
On this time-scale we can make the standard rotating 
wave approximation (RWA), this being terms oscillating 
at frequencies greater than or equal to $\omega_{0}$ 
take their average value (unless they multiply white 
noise terms, that have non-negligible components at 
$\omega_0$).  

Applying the above to Eqs. (\ref{eqs:xydot}) 
%(\ref{eq:xdot}) and (\ref{eq:ydot}) 
gives 
 \begin{subequations}\begin{align}
 \rmd \tilde{x}(t) &= 
   \sq{-\frac{\gamma}{2}\tilde{x}(t)
   -\ortega_\mathrm{x}V_\mathrm{in}^{\rf}
   +\ortega_\mathrm{y}
%   {\red {\sqrt{S_\mathrm{{ \tilde{I}}}^{\sin}}}}
         {{ \tilde{I}}^{\sin}(t)}
      }\dt ,
 \label{eq:dxtilde:2}
 \\
 \rmd \tilde{y}(t) &= 
   \sq{-\frac{\gamma}{2}\tilde{y}(t)
   -\ortega_\mathrm{y}
%   {\red {\sqrt{S_\mathrm{{ \tilde{I}}}^{\cos}}}}
         {{ \tilde{I}}^{\cos}(t)}
      }\dt ,
 \label{eq:dytilde:2}
 \end{align}\label{eqs:dxdytilde:2}\end{subequations}
where the tilde denotes that we are in the frame rotating at 
$\omega_0$ and we have dropped the effect of the input noise, $N(t)$,  as we assume that it is mainly $1/f$ noise 
and has negligible spectral weight at $\omega_0$. The two currents 
${ \tilde{I}}^\mathrm{cos}(t)$ and ${\tilde{I}}^\mathrm{sin}(t)$ are the 
two quadratures of the filtered detector current ${\tilde{I}}(t)$ 
centered at the frequency $\omega_0$.  That is, they are given by
\begin{subequations}\begin{align}
%{\red \sqrt{S_\mathrm{{\tilde{I}}}^{\cos}}}
{\tilde{I}}^\mathrm{cos}(t) &\equiv 
\frac{1}{\dt} \int_{t-\dt}^{t}{\tilde{I}}(s)\cos(\omega_{0}s)\rmd s , 
\label{eq:Jx}\\
%{\red \sqrt{S_\mathrm{{\tilde{I}}}^{\sin}}}
{\tilde{I}}^\mathrm{sin}(t) &\equiv 
\frac{1}{\dt} \int_{t-\dt}^{t}{\tilde{I}}(s)\sin(\omega_{0}s)\rmd s .
\label{eq:Jy}
\end{align}\label{eqs:JxJy}\end{subequations}
These components are band-pass-filtered currents over 
the bandwidth $\rmd t^{-1}$ and are the quantities to 
which an idealized observer would have access.
To be more specific, the current coming from the QPC is 
\begin{equation}
 {\tilde{I}} (t) = e |{\cal T}| J(t) +{\tilde{I}}_\mathrm{clas}(t) ,
\label{eq:specific}
\end{equation}
where $J(t)$ is the quantum signal given by 
\begin{equation}
	J(t) = \sqrt{\kappa_0(t)}\langle \hat \sigma_z\rangle + \xi(t).
\end{equation} 
{where the rate $\kappa_0$ will be discussed later, and 
$\xi(t)$ is the QPC shot noise, which we take to be white in the 
limit of a weakly responding QPC.\cite{KorPRB99,KorPRB01b}  
That is, the QPC shot noise is Gaussian noise with the following 
correlations 
\begin{equation}
	\begin{split}
		\mathrm{E}[\xi(t)] &= 0\\
		\mathrm{E}[\xi(t)\xi(t')] &= \delta(t-t').
	\end{split}
\label{eq:EdW}
\end{equation} 
Here $\mathrm{E}$ denotes an ensemble average over all possible 
realizations of $\xi(t)$.  
In \Eq{eq:specific} ${\tilde{I}}_\mathrm{clas}(t)$ is the 
deterministic classical component of the current (with the 
steady-state current subtracted --- see Appendix 
\ref{app:recastSME} for details).  Using \Eq{eq:Vd} we can 
write 
\begin{align}
     |{\cal T (t)}| &\approx | {{\cal T}_0}|
	 \sq{1 + \half\epsilon_\mathrm{in}\sin\ro{\omega_{0}t}} ,
\label{eq:T(t)}
\end{align}
since $\epsilon_\mathrm{in}\ll 1$ (see \Sec{sec:Vd}).  
Using this, the two %{\red normalized} 
quadrature currents become 
\begin{subequations}\begin{align}
\label{eq:Icos} 
%{\red \sqrt{S_\mathrm{{\tilde{I}}}^{\cos}}}
{\tilde{I}}^\mathrm{cos}(t)  &= e |{{\cal T}_0}| 
{ \sqrt{S^{\cos}}}
J^{\mathrm{cos}}(t)
 + {\tilde{I}}^{\mathrm{cos}}_\mathrm{clas}(t) , 
\\ 
\label{eq:Isin}
%{\red \sqrt{S_\mathrm{{\tilde{I}}}^{\sin}}}
{\tilde{I}}^\mathrm{sin}(t)  &= e |{{\cal T}_0}| 
{ \sqrt{S^{\sin}}}
J^{\mathrm{sin}}(t) + {\tilde{I}}^{\mathrm{sin}}_\mathrm{clas}(t) ,
\end{align}\label{eqs:Icossin}\end{subequations}
where $J^{\mathrm{cos}}(t)$ and 
      $J^{\mathrm{sin}}(t)$ are the 
two quadrature components of $J(t)$.  
They are 
\begin{subequations}\begin{align}
\label{eq:Jidealcos}
J^{\mathrm{cos}}(t) &=\frac{1}{\sqrt{S^{\cos}}}\left[
J^{\mathrm{cos}}_{\omega_0}(t) 
+ \frac{1}{4}\epsilon_\mathrm{in} 
J^{\mathrm{sin}}_{2\omega_0}(t) \right],
 \\
\label{eq:Jidealsin}
J^{\mathrm{sin}}(t) &=
\frac{1}{\sqrt{S^{\sin}}}\left[J^{\mathrm{sin}}_{\omega_0}(t) 
+ \frac{1}{4}\epsilon_\mathrm{in}
[J_{0}^{\mathrm{cos}}(t) 
 -J^{\mathrm{cos}}_{2\omega_0}(t)]\right] , 
\end{align}\label{eqs:Jideal}\end{subequations}
 where $S^{\cos}$ and $S^{\sin}$ 
are dimensionless and defined such that 
$[J^{\mathrm{cos}}(t)\dt]^2 =[J^{\mathrm{sin}}(t)\dt]^2 = \dt$.  
The Fourier components of the quantum 
signal $J(t)\dt$ are 
\begin{subequations}\begin{align}
\label{eq:Jcos}
J^{\mathrm{cos}}_\mathrm{n \omega_0}(t) &=
\frac{1}{\dt} \int_{t-\dt}^{t} J(s)\cos(n \omega_{0} s) \rmd s ,
\\
\label{eq:Jsin}
J^{\mathrm{sin}}_\mathrm{n \omega_0}(t) &=
\frac{1}{\dt} \int_{t-\dt}^{t}J(s) \sin(n \omega_{0} s) \rmd s .
\end{align}\label{eqs:Jcossin}\end{subequations}
Note that 
$[J^{\mathrm{sin}}_\mathrm{n\omega_0}(t)\dt]^{2}
=[J^{\mathrm{cos}}_\mathrm{n\omega_0}(t)\dt]^{2}
=\dt/2$ for $n=1,2$; and that 
$[J^{\mathrm{cos}}_\mathrm{0}(t)\dt]^{2}
=\dt$ {[using \Eq{eq:EdW}]}. 

To measure the two quadrature currents in \Eq{eqs:Icossin} 
[and hence the quantum signals 
$J^{\mathrm{cos}}(t)$ and 
$J^{\mathrm{sin}}(t)$], 
an ideal observer would measure the desired quadrature of the 
output voltage.  This can be done by beating the output 
voltage with a local oscillator.\cite{Pozar98}  
Using \Eq{eq:vout}, the two output voltage quadratures are 
\begin{subequations}\begin{align}
V_{\mathrm{out}}^{\mathrm{cos} }(t) 
& = 
\frac{ B_{\sin} }{\ortega_\mathrm{x}}\frac{\gamma}{2} 
\int_{0}^{t}
J^{\mathrm{sin}}(s) 
            e^{-\gamma (t-s)/2}ds ,  
\label{eq:Voutx}\\
V_{\mathrm{out}}^{\mathrm{sin} }(t)
&= 
-\frac{B_{\cos} }{\ortega_\mathrm{x}} \frac{\gamma}{2}
\int_{0}^{t} 
J^{\mathrm{cos}}(s) 
e^{-\gamma (t-s)/2}ds ,
\label{eq:Vouty}
\end{align}\label{eqs:convolutions}\end{subequations} 
where $B_{\sin} = e|{\cal T}_0|\ortega_\mathrm{y} 
\sqrt{S^{\sin}}$, and 
$B_{\cos} = e|{\cal T}_0|\ortega_\mathrm{y} \sqrt{S^{\cos}}$ have units of $t^{-1/2}$.  
That is, $B_{\sin}^2$ is the proportionality constant that 
turns the quantum signal, $J(t)$, into the dimensionless current 
$\alpha_\mathrm{y} {\tilde{I}}^{\sin}$, which drives 
the classical circuit (the same is also true for the 
cosine quadrature component).  
Here we have done some post-processing, removing the 
deterministic parts of $V_{\mathrm{out}}^{\cos}(t)$ 
and $V_{\mathrm{out}}^{\sin}(t)$, and dropping 
all contributions from $N(t)$ as again we assume that 
it has negligible spectral weight at $\omega_0$.  
By inverting these convolutions, the idealized observer 
can access both $J^{\cos}(t)$ and 
$J^{\sin}(t)$, enabling them to 
condition the quantum state on these currents.  The 
resulting equation describing the idealized conditional 
dynamics of the qubit is called a stochastic master 
equation, or quantum trajectory equation.  
Note that we can write \Eq{eq:Voutx} as 
$V_{\mathrm{out}}^{\cos}(t) 
 = \gamma\tilde{x}(t)/(2\ortega_\mathrm{x})$ 
(after some post processing), and similarly for 
\Eq{eq:Vouty} in terms of $\tilde{y}$.

%****************************************************************
\section{Idealized quantum dynamics
\label{sec:qubit}}
We now consider the idealized case where the stochastic 
QPC current can be perfectly measured [we have access 
to both $J^{\mathrm{cos}}(t)$ and $J^{\mathrm{sin}}(t)$].  
To describe the idealized conditional qubit dynamics, 
we start with the microscopic model of charge qubit 
monitoring by a (dc) QPC in \Ref{GoaMilWisSunPRB01}.  
We make an important modification to the model that is 
due to the time-dependence of the voltage across 
the rf-QPC, $V_{\rmd}(t)$ [see \Eq{eq:Vd}].  
It results in time-dependent QPC tunnelling rates.  
We also make a rotating wave 
approximation (RWA) to simplify the analysis.  
The RWA is only possible for a weakly responding QPC.  
In this paper, as in \Ref{GoaMilWisSunPRB01}, we will refer to 
the {limit of weak response} as quantum diffusion.  
This is because in this limit there are many electrons passing 
through the QPC with each containing only a little information 
about the qubit state, which under monitoring of the QPC current 
the evolution of the qubit will slowly wander towards one of the 
$\sig_z$ eigenstates rather then a sudden collapse.  
The linear form of the qubit stochastic master equation of 
\Ref{GoaMilWisSunPRB01} is (see Appendix \ref{app:recastSME} 
for details)
\begin{align}
d\bar{\rho}_\mathrm{c}(t) 
&\equiv 
- {\frac{i}{\hbar}}
\sq{\Ham_\mathrm{qb}^{\prime}(t)+
\hat{H}_\mathrm{J}(t),\bar{\rho}_\mathrm{c}(t)} dt
\nn\\
&\phantom{=}
+ \half \Gamma_{\rmd}(t)
\D{\sig_{z}}\bar{\rho}_\mathrm{c}(t) dt
\nn \\ &\phantom{=}
+
[J(t)-\mu]dt%\frac{J(t) dt}{\sqrt{S^\mathrm{J}(t)}} 
%\cu{
\hlin{\sqrt{\kappa_{0}(t)}\sig_{z}/2}
%      }
\bar{\rho}_\mathrm{c}(t) ,
\label{eq:sme} 
\end{align}
where 
\begin{subequations}\begin{align}
\Ham_\mathrm{qb}^{\prime}(t)&=
\Ham_\mathrm{qb}+ {\hbar}\sig_{z}\st{\Tcal(t)}\st{\Xcal(t)}
\sin(\theta)/2 , 
\label{eq:Hqbprime}
\\
\hat{\bar{H}}_\mathrm{J}(t) &=
-{\hbar}\sig_{z}[J(t)-\mu] \sqrt{\kappa_{1}(t)}/2 ,
\label{eq:HJ}
\end{align}\label{eqs:H}\end{subequations}
and $\mu$ is the mean of the ostensible 
distribution used to derive \Eq{eq:sme} 
(see Appendix \ref{app:recastSME}).  
The linear superoperator, $\bar{\cal H}$, in \Eq{eq:sme} 
is defined for arbitrary operators $\hat{c}$ by 
 \begin{equation}
 \hlin{\hat{c}}\rho = \hat{c}\rho + \rho\hat{c}\dg -\mu\rho.
 \label{eq:Hlinear}
\end{equation}
It represents the back action effects of the continuous measurement.  
The time-dependent qubit dephasing rate is 
$\Gamma_{\rmd}(t)=[\kappa_{0}(t) + \kappa_{1}(t)]/2$, 
where $\kappa_{0}(t)$ and $\kappa_{1}(t)$ represent two types 
of measurement-induced dephasing in the qubit: 
$\kappa_{0}(t)$ represents 
%``pure'' dephasing, also called 
information-limited (Heisenberg) dephasing,\cite{KorPRB03} 
which reflects the quantum-mechanical cost of obtaining 
information about the qubit state; 
$\kappa_{1}(t)$ represents additional (non-Heisenberg) 
%``impure'' 
dephasing by processes that yield no qubit information.  
We define these dephasing rates by
\begin{subequations}\begin{align}
\sqrt{\kappa_{0}(t)}
&\equiv 
\st{\Xcal(t)}\cos\ro{\theta} 
= \sqrt{2\Gamma_{\rmd}(t)}\cos\ro{\theta} ,
\label{eq:kappa0}
\\
\sqrt{\kappa_{1}(t)}
&\equiv \st{\Xcal(t)}\sin\ro{\theta} 
= \sqrt{2\Gamma_{\rmd}(t)}\sin\ro{\theta} ,
\label{eq:kappa1}
\end{align}\label{eqs:kappas}\end{subequations}
where $\theta$ is the relative phase between the QPC 
tunnelling amplitudes ${\cal T}$ and ${\cal X}$.  
Note that if $\theta=0$ or $\theta=\pi$, then 
$\kappa_{1}=0$, and $\hat{\bar{H}}_\mathrm{J}(t)=0$.  
This is a necessary condition for the QPC to be considered 
an ideal charge qubit detector operating at the quantum 
limit.\cite{KorPRB99,AveSukPRL05,OxtWisSunPRB06}  

From \Eq{eq:Vd} we can write
\begin{align}
\st{\Xcal(t)} &\approx \sqrt{\kappa}
\sq{1 + \half\epsilon_\mathrm{in}\sin\ro{\omega_{0}t}} ,
\label{eq:Gammad}
\end{align}
since $\epsilon_\mathrm{in}\ll 1$ (see \Sec{sec:Vd}).  
Here the time-dependence has been made explicit, and 
{ $\kappa/2 = \st{\mathcal{X}}^2 / 2$ }
is the time-independent dephasing rate for the charge 
qubit monitored by a dc-QPC (the situation in 
\Ref{GoaMilWisSunPRB01}).  
Substituting $\kappa_{0}(t)$ and $\kappa_{1}(t)$ (with the 
above approximation for their time dependence) into 
\Eq{eq:sme} {and making the RWA} gives 
\begin{align}
\rmd {\bar{\rho}}_\mathrm{c}(t)
&=
- {\frac{i}{\hbar}}
   \sq{{ \widetilde{H}_\mathrm{qb}}
      +\widetilde{H}_\mathrm{J}(t),
       {\bar{\rho}}_\mathrm{c}(t)} \dt
\nn\\
&\phantom{=}
+ \kappa \D{\sig_{z}}\bar{\rho}_\mathrm{c}(t) \dt /{ 4}
\nn \\
&\phantom{=}
+ \cu{{J}^{\mathrm{cos}}_ 0(t)-\mu
     +\frac{\epsilon_\mathrm{in}}{2}
      \sq{{J}^{\mathrm{sin}}_{\omega_0}(t)-\mu}}\dt
	  \nn \\
&\phantom{=+}
\times
\hlin{\sqrt{\kappa}\cos(\theta)\sig_{z}/2}
{\bar{\rho}}_\mathrm{c}(t) ,
\label{eq:QPCsmeRWA}
\end{align}
where ${J}^{\mathrm{cos}}_0(t)$ and ${J}^{\mathrm{sin}}_{\omega_0}(t)$ are defined in 
Eqs. (\ref{eqs:Jcossin}).  
The rotated versions of the Hamiltonians are 
\begin{subequations}
\begin{align}
\widetilde{H}_\mathrm{J}(t) = &
-{\hbar}\sig_{z}\cu{{J}_{0}^\mathrm{cos}(t)-\mu
    +\frac{\epsilon_\mathrm{in}}{2} \sq{{J}_{\omega_0}^\mathrm{sin}(t)-\mu}
    }\nn
\\&\times\sqrt{\kappa}\sin(\theta)/2 ,
\label{eq:HJtilde}\\
\widetilde{H}_\mathrm{qb} \approx &
\Ham_\mathrm{qb}
+ {\hbar}\sig_{z}\st{\Tcal_{0}}\sqrt{\kappa}
  \sin(\theta)/2 .
\label{eq:Hqbtilde}
\end{align}\label{eqs:Hrotated}\end{subequations}

To get the linear form of the ideal quantum trajectory 
we need to rewrite \Eq{eq:QPCsmeRWA} in terms of the 
signal that an ideal observer could access, namely 
$J^\mathrm{sin}(t)$.  This is achieved by 
expressing $J^{\mathrm{cos}}_0(t)$ and 
$J^{\mathrm{sin}}_{\omega_0}(t)$ in terms of 
both the observed process, $J^\mathrm{sin}(t)$, 
plus some other unobserved processes, as done in 
\Ref{OxtetalPRB05}.  Averaging over the unobserved 
processes results in the following 
linear quantum trajectory equation in the RWA: 
\begin{align}
\rmd {\bar\rho}_\mathrm{c}(t)
&= 
\tilde{\mathcal{L}}{\bar\rho}_\mathrm{c}(t)\dt
\nn\\
&\phantom{=}
+ \sqrt{\eta}\sq{J^\mathrm{sin}(t)
{- \sqrt{\eta}\mu}}\dt 
\hlin{\sqrt{\kappa}\sig_{z}/2}{\bar\rho}_\mathrm{c}(t) ,
\label{eq:QPCsmeRWA:linear}
\end{align}
where we have defined the efficiency $\eta$ by 
\begin{equation}
\eta \equiv 
\epsilon^2_\mathrm{in} 
\cos^2(\theta) 
{S^\mathrm{sin}} \approx \epsilon^2_\mathrm{in} \cos^2(\theta)/4 ,
\label{eq:eta}
\end{equation}
and the Liouvillian $\tilde{\cal L}$ is 
\begin{align}
\tilde{\cal L} \rho= - {\frac{i}{\hbar}}
\sq{\widetilde{H}_\mathrm{qb}
+\widetilde{H}_\mathrm{J}^{\prime}(t),
       {\rho}} + \frac{\kappa}{4} \D{\sig_{z}}{\rho} . 
\label{eq:Ltilde}
\end{align}
We now have 
$\widetilde{H}_\mathrm{J}^{\prime}(t) = 
-\hbar\sig_{z}J^{\sin}(t)
\sqrt{\eta\kappa}
\tan(\theta)/2$.  
Equation (\ref{eq:QPCsmeRWA:linear}) {normalizes to} 
\begin{align}
 \rmd {\rho}_\mathrm{c}(t)
 &=
\tilde{\cal L}{\rho}_\mathrm{c}\dt
+ 
\sqrt{\eta}
    \sq{
     J^\mathrm{sin}(t)
          -\sqrt{\eta\kappa}
        \an{\sig_{z}}}\dt 
\nn\\&\phantom{=++++} 
\times 
\h{\sqrt{\kappa}\sig_{z}/2}{\rho}_\mathrm{c}(t) ,
\label{eq:QPCsmeRWA:normed}
\end{align}
where the normalized current is 
\begin{equation}
	J^\mathrm{sin}(t)\dt=\sqrt{\eta\kappa}
	\an{\sig_{z}}\dt+ \rmd W(t)
\end{equation} 
{where $\rmd W(t) = \xi(t) \rmd t$ is a Wiener 
increment.\cite{QNoise}  }
This is the explicit expression for the quantum signal to 
which an ideal observer would have direct access.  
To get the correct statistics, we have made a Girsanov 
transformation,\cite{GatGisJMP91} which results in 
replacing $\mu$ in the above with 
$\sqrt{\kappa}\an{\sig_{z}}$.  
The stochastic Hamiltonian 
$\widetilde{H}_\mathrm{J}^{\prime}(t)$ in \Eq{eq:Ltilde} 
will also be updated in the same manner (replacing $\mu$ with 
$\sqrt{\kappa}\an{\sig_{z}}$).  
The nonlinear back action superoperator is defined by 
its action on $\rho$ with an arbitrary operator $\hat{c}$ 
by
\begin{equation}
{\cal H}[\hat{c}]\rho = \hat{c} \rho +\rho \hat{c}\dg 
- \an{\hat{c}+\hat{c}\dg}\rho .
\label{eq:Hnonlinear}
\end{equation}

Equation (\ref{eq:QPCsmeRWA:normed}) is the first 
stochastic master equation presented for continuous 
measurement of a charge qubit using an rf configuration.  
It reveals some interesting {\label{discussion:reveals}}physics 
about the rf-QPC operating in the rf+dc mode.  
First, we note that the qubit dephasing is due only to the 
dc component of the detector voltage, since $\kappa$ is 
a function of $V_{0}$ [and is independent of the AC 
component of $V_{\rmd}(t)$].  
Second, the rf-QPC is a highly inefficient detector when 
operated in the rf+dc mode since 
$\eta \propto \epsilon_\mathrm{in}^{2}\ll1$.  
Physically, this low detection efficiency arises 
because the qubit information is extracted relatively 
slowly by the (small) rf component of the QPC voltage, 
compared to the qubit dephasing by the (large) dc 
component.  
We conjecture that the low measurement efficiency of the 
rf+dc QPC may in practice be higher than that of the dc-QPC 
for two reasons: $1/f$ noise limits the dc-QPC measurement 
efficiency and also further degrades the efficiency 
the longer the measurement has to continue.

%*************************************************************
\section{Realistic dynamics: realistic quantum trajectory 
equation \label{sec:probability}}
The previous sections conditioned the qubit evolution 
and the classical oscillator evolution on idealized 
measurement results available only to a hypothetical 
observer.  For an experimentalist, it is much more 
useful to consider how to describe the qubit evolution 
conditioned on measurement results available in the 
laboratory.  
In rf-QPC or rf-SET experiments, the voltage leaving the 
transmission line is observed using homodyne detection 
of the amplitude quadrature ($\tilde{x}$).  
The phase quadrature ($\tilde{y}$) can be ignored 
because it is independent of the qubit-information-carrying 
signal, $J^{\sin}(t)$ 
[see Eqs. (\ref{eqs:dxdytilde:2}) and (\ref{eqs:convolutions})].  

In simple dyne detection 
(see schematic in \Fig{fig:rfcircuit}), 
the output signal 
$V_\mathrm{out}(t)$ is amplified, 
and mixed with a local oscillator (LO).  
The LO for homodyne detection of the amplitude 
quadrature is 
$V_\mathrm{LO}(t) \propto \cos\ro{\omega_{0}t}$, 
where the LO frequency is 
the same as the signal of interest (or very slightly 
detuned).  
{The resulting low-frequency beats due to mixing the 
signal with the LO are easily detected.}  

The signal resulting from the homodyne detection is 
\begin{align}
\V{}(t) \dt
&=
\sqrt{A_\mathrm{SN}}
\frac{\ortega_\mathrm{x}}{B_{\mathrm{sin}}} 
V^{\cos}_\mathrm{out}(\tilde{x})\dt
+\rmd W_\mathrm{out}(t) %.
\nn%\label{eq:Vcal:x:1}
\\
&\equiv 
\lambda_{\tilde{x}} \dt 
+\rmd W_\mathrm{out}(t) ,
\label{eq:Vcal}
\end{align}
where $A_\mathrm{SN}$ is the ratio of gain squared 
to noise power which has the dimensions of inverse time.  This relates to 
the dimensionless signal-to-noise ratio for the measurement as 
$\mathrm{SNR} = A_\mathrm{SN}t$ ($t$ is the measurement duration).  
Equation (\ref{eq:Vcal}) shows that white noise is added to the 
amplified quadrature signal.  Here 
\begin{equation}
\lambda_{\tilde{x}} \equiv 
\sqrt{A_\mathrm{SN}}
\frac{\gamma\tilde{x}}
{2B_\mathrm{sin}} ,
\label{eq:lambda}
\end{equation}
and the output noise Wiener increment 
$\rmd W_\mathrm{out}(t)$ satisfies 
$[\rmd W_\mathrm{out}(t)]^{2}=\dt$.  
The addition of this extra noise prevents one from inverting 
the convolutions in Eqs. (\ref{eqs:convolutions}) to find 
$J^{\sin}(t)$, and so the realistic 
quantum trajectory approach\cite{WarWisMabPRA02,WarWisJOBQSO03a} 
must be employed to condition the qubit evolution on 
the realistic measurement record $\V{}(t)$.  

Our derivation of the realistic quantum trajectory equation 
describing the conditional evolution of the combined 
oscillator-qubit state closely follows the standard 
techniques.\cite{WarWisJOBQSO03a,OxtetalPRB05}  The details 
are included in Appendix \ref{app:RQTderivation}, with the 
final result being the following superoperator 
Kushner-Stratonovich (SKS) equation: 
\begin{align}
\rmd {\rho}_{\V{}}(\tilde{x}) 
 &= 
 \Bigg\{
   \sq{-\frac{\du}{\du\tilde{x}}
           m_{\tilde{x}}
     +\frac{B_{\sin}^2}{2}
      \frac{\du^{2}}{\du\tilde{x}^{2}}
     +\tilde{\mathcal{L}}}\dt
 \Bigg.\nn\\
 &\phantom{==}
 \Bigg.
  + %   \frac{\lambda_{\tilde{x}}-\lambda_{\an{\tilde{x}}_{\rho}}}
  (\lambda_{\tilde{x}}-\lambda_{\an{\tilde{x}}_{\rho}})[\V{}(t) - \lambda_{\an{\tilde{x}}_{\rho}}]\rmd t
\Bigg\} 
 {\rho}_{\V{}}(\tilde{x})
\nn\\
 &\phantom{==}
 -\dt\sqrt{\eta\kappa}B_{\sin}\frac{\du}{\du\tilde{x}}
%   \frac{\epsilon_\mathrm{in}}{4}\sqrt{\kappa}\cos(\theta)
\left[\sig_{z}{\rho}_{\V{}}(\tilde{x})+
{\rho}_{\V{}}(\tilde{x})\sig_{z}\right]/{ 2} ,
\label{eq:skse:rfqpc}
\end{align}
{where we now consider $\tilde{x}$ to be 
post-processed (the deterministic part has been removed), 
and we have defined $m_{\tilde{x}}\equiv-\gamma\tilde{x}/2$.}  
The true record the realistic observer would measure is 
\begin{align}
\V(t)\dt = \lambda_{\an{\tilde{x}}_{\rho}}\dt + \rmd \W (t) ,
\label{eq:Vtrue}
\end{align}
{ %
where the observed white noise $\rmd\W(t)$ is 
not the same as $\rmd W_\mathrm{out}(t)$ 
[contrast (\ref{eq:Vtrue}) and (\ref{eq:Vcal})].}  
In \Eq{eq:skse:rfqpc}, the Liouvillian superoperator $\mathcal{L}$ 
contains the Hamiltonian evolution and qubit dephasing (and is 
defined in Appendix \ref{app:RQTderivation}).  
The realistic quantum trajectory equation (\ref{eq:skse:rfqpc}) 
is an important result of this paper.  
The first line describes the uncoupled, average 
evolution of the oscillator and the qubit.  The second 
line describes the update of the realistic observer's 
state of knowledge of the supersystem, conditioned 
on the realistic homodyne output.  The final line 
contains the effect of the qubit on the circuit.  

The (reduced) state of the qubit conditioned on 
the realistic output, $\rho_{\V{}}$, is found from 
$\rho_{\V{}}(\tilde{x})$ by integrating out the circuit variable 
$\tilde{x}$.  Similarly, the (marginal) state of the circuit can 
be found by tracing out the qubit.

%**************************************************************** 
\section{Oscillator slaved to qubit 
\label{sec:slaved}}
For sufficiently large damping $\gamma$ such that the first 
term in \Eq{eq:dxtilde:2} dominates the dynamics of $\tilde{x}$ 
{(but still $\gamma_{\rmd}\gg\gamma$, so that {the 
RWA remains valid})}, the oscillator immediately damps to a 
qubit-dependent state.  
The circuit is thus said to be \textit{slaved} to the qubit.  
This adiabatic elimination of the circuit allows the conditioned 
qubit dynamics to be once again governed by the stochastic master 
equation alone.  
However, a realistic observer has access only to the 
homodyne output, that involves excess noise above that of 
$J^\mathrm{sin}$.  
Thus, we find the slaved output 
to which the realistic observer has access, then use it to 
condition the qubit state.  
The slaved value of $\tilde{x}$ is found by taking 
$\gamma\rightarrow\infty$ in \Eq{eq:Voutx}.  
This allows us to make the standard delta-function 
replacement and write the quadrature output voltage as 
$V_{\mathrm{out}}^\mathrm{cos}(t) = B_{\mathrm{sin}} 
J^{\mathrm{sin}}(t)/\ortega_\mathrm{x}$.  
That is, the quadrature output voltage is directly 
proportional to the quantum signal an ideal observer 
would measure.  Substituting this into \Eq{eq:Vcal} 
gives 
\begin{align}
\V{}_\mathrm{sl}(t)\dt = \sqrt{A_\mathrm{SN}} 
J^{\mathrm{sin}}(t) \dt
+  \rmd W_{\mathrm{out}}(t),
\label{eq:Vslaved}
\end{align} 
where $J^{\mathrm{sin}}(t)$ is the 
signal used to condition the qubit state in 
\Eq{eq:QPCsmeRWA:linear}.  
Here we see that the output noise degrades the QPC 
signal, the effect of which is to reduce the efficiency 
of the detection. 

Using the above we can redefine the quantum signal a realistic observer 
would measure as 
$\V{}_\mathrm{sl}(t)= J_\mathrm{sl}(t) 
\sqrt{A_\mathrm{SN}} + 1$, 
where
\begin{align}
J_\mathrm{sl}(t) \dt = \sqrt{\eta_\mathrm{sl}\kappa} \an{\sig_z} \dt+ \rmd {\cal W}(t),
\end{align} where 
\begin{align} 
\label{eq:rfqpc:sme:slaved}
\eta_\mathrm{sl} = \eta \frac{A_\mathrm{SN}}
{A_\mathrm{SN} + 1} 
\end{align} 
is the efficiency of the realistic rf+dc QPC in the slaved 
limit.  
Using this quantum signal, the stochastic master equation in the 
slaved limit is 
\begin{align}
\rmd {\rho}_\mathrm{c}(t) &=
\dt \tilde{\cal L}{\rho}_\mathrm{c}(t) + \sqrt{\eta_\mathrm{sl}}\sq{ J_\mathrm{sl}(t)- \sqrt{\eta_\mathrm{sl}
\kappa}  \an{\sig_{z}}}\dt\nn\\
&\phantom{=}
\qquad\qquad\qquad
\times\h{\sqrt{\kappa}\sig_{z}/2}{\rho}_\mathrm{c}(t) ,
\label{eq:QPCsmeRWA:slaved}
\end{align}
where the noise term in $\tilde{\cal L}$ is now in terms 
of $J_\mathrm{sl}(t)$.  
The fact that the efficiency $\eta_\mathrm{sl}$ in 
\Eq{eq:rfqpc:sme:slaved} is strictly less than unity shows 
that the rf+dc QPC is {incapable} of reaching the quantum 
limit of a purity-preserving detector, even in the slaved 
limit.  
This conclusion is expected from the 
discussion following the idealized stochastic master 
equation in the RWA, \Eq{eq:QPCsmeRWA:normed}.  
That is, even without including the circuit in our description 
of the rf-QPC, \Eq{eq:QPCsmeRWA:normed} shows that, for 
$\epsilon_\mathrm{in}\ll1$ (the rf+dc mode), the rf-QPC is a 
highly inefficient charge qubit detector (that may nonetheless 
be more efficient than the dc-QPC due to the effects of 
$1/f$ noise in the dc case).  
Note that the theoretical maximum of $\eta_\mathrm{sl}$ is 
the idealized efficiency $\eta$, which occurs for 
infinite signal-to-noise ratio in the amplifier.  
As a final point, we note that the inclusion of input
white noise in the analysis will have the effect of further
reducing the rf+dc QPC efficiency (and significantly
complicating the analysis).\cite{OxtPHD07}

%************************************************************
\section{Discussion and summary
\label{sec:conclusion}}
In this paper we have presented a model for conditional 
monitoring of a charge qubit using an rf 
(radio-frequency) configuration.  The rf configuration was 
introduced for the single-electron transistor (SET) in 
\Ref{SchetalSC98}.  
It involves embedding the SET in a resonant (oscillator) 
circuit, and monitoring the resulting damping.  
This allows operation of the SET at high frequencies, 
where the $1/f$ noise prevalent in conventional 
low-frequency, dc measurements is completely negligible.  

{We have two main results.  We have derived an evolution 
equation for the conditional state of a charge qubit monitored 
continuously in time by a detector operated in the rf 
configuration.  To the best of our knowledge, an equation of 
this type has not previously been derived.  
This culminates in the stochastic master equation, or quantum 
trajectory, \Eq{eq:QPCsmeRWA:normed}.  
Our second main result is the extension of our idealized 
quantum trajectory (\ref{eq:QPCsmeRWA:normed}) to consider 
conditioning the qubit state on a corrupted (filtered, more 
noisy) measurement signal available to a realistic observer.  
Our realistic quantum trajectory\cite{WarWisMabPRA02,WarWisJOBQSO03a,OxtetalPRB05} 
equation is \Eq{eq:skse:rfqpc}.  
}

Our model used the quantum-point contact (QPC) in 
place of the SET, and also assumed operation of the 
rf-QPC in the so-called\cite{AasetalAPL01} rf+dc mode.  
In this mode, a small amplitude sinusoid (carrier) at the 
resonance frequency of the oscillator is superposed on a 
relatively large bias (dc).  
{Experimentally\cite{AasetalAPL01} this is done 
to maximize the sensitivity of a SET.  In this mode we 
can make a rotating wave approximation (RWA) in the limit 
of a vanishingly small amplitude sinusoid.  
We found the rf+dc QPC to be a highly inefficient 
charge-qubit detector, even ignoring the external circuit.  
Physically, this is because the small rf amplitude 
carries the qubit information, while the relatively large 
dc bias causes most of the measurement-induced qubit 
dephasing.  
It is important to note that the low measurement efficiency 
of the rf+dc QPC may in practice be higher than for the 
dc-QPC due to the debilitating effects of $1/f$ noise.  
Specifically, $1/f$ noise limits the dc-QPC measurement 
efficiency, and also further degrades the efficiency 
the longer the measurement has to continue.
}

Having realistically modeled continuous-in-time qubit 
measurement using the rf+dc QPC, the next step is to 
extend realistic quantum trajectory theory to the pure 
rf mode (zero bias, larger rf amplitude) for both the QPC 
and the SET.  These are important future tasks, particularly 
for understanding and designing measurement and feedback 
control of quantum systems in the solid-state --- valuable 
knowledge for fully harnessing the potential of future 
quantum technologies such as quantum computers.

\acknowledgments
This work was supported by the ARC and the State of Queensland. 
NO acknowledges partial support from the QIP IRC and QAP.  
JG was partially supported by MITACS and ORDCF.

\appendix
\section{Recasting the stochastic master equation of 
Goan, \etal \label{app:recastSME}}
The microscopic model of charge qubit monitoring by a 
conventional (dc) QPC in \Ref{GoaMilWisSunPRB01} is used 
in this appendix as a starting point for producing a 
stochastic master equation for the rf-QPC (in the 
rf+dc mode) [Eq. \ref{eq:sme}].  We add time dependence to the QPC 
tunnelling rates that is due to the time-dependent 
rf-QPC voltage (see \Sec{sec:Vd}), and recast the 
diffusive stochastic master equation of 
\Ref{GoaMilWisSunPRB01} into a linear form.

The diffusive stochastic master equation of 
\Ref{GoaMilWisSunPRB01} is 
%\begin{align}
%d\rho_\mathrm{c}(t) &= 
%-i\sq{\Ham_\mathrm{qb},\rho_\mathrm{c}(t)}
%+\D{{\Tcal} + \Xcal\hat{n}}\rho_\mathrm{c}(t) \nn\\
%&\phantom{=}
%+\highlight{\xi(t)}\frac{}
%\end{align}
\begin{align}
d\rhoc(t) &= 
- {\frac{i}{\hbar}}
\sq{\hat{H}_\mathrm{qb},\rhoc(t)} dt
+ \D{\mathcal{T} + \mathcal{X}\hat{n}}\rhoc(t) dt
\nlalign{
+ \xi(t)dt\frac{1}{\st{\mathcal{T}}}
\sq{\mathcal{T}^{*}\mathcal{X}\hat{n}\rhoc(t)
  + \mathcal{X}^{*}\mathcal{T}\rhoc(t)\hat{n}
   }
}
\nlalign{
- \xi(t)dt\frac{1}{\st{\mathcal{T}}}
2\mathrm{Re}\ro{\mathcal{T}^{*}\mathcal{X}}
   \an{\hat{n}}_\mathrm{c}\rhoc(t)
}
\label{app:eq:QPCsme1}
\end{align}
where $\hat{n} = \ket{1} \bra{1}$ and $\hat{H}_\mathrm{qb}$, 
and the time-dependencies of $\Tcal(t)$ and $\Xcal(t)$ in our 
model are defined in \Sec{sec:quantum}.  
This equation can be rewritten as 
 \begin{align}
 d\rhoc(t) &= 
 - {\frac{i}{\hbar}}
 \sq{\hat{H}^{\prime}_\mathrm{qb}(t),\rhoc(t)} dt
 + \half \Gamma_{\rmd}(t)\D{\sig_{z}}\rhoc(t) dt
\nn\\
&\phantom{=}
	 %\mathcal{L}\rhoc(t) dt
	 + \xi(t)dt\left\{
	 \h{\sqrt{\kappa_{0}(t)}\sig_{z}/2}
	 \right.
\nn\\
&\phantom{=+ \xi(t)dt}
	 \left.	 
     +\h{\sqrt{\kappa_{1}(t)}i\sig_{z}/2}\right\}\rhoc(t),
\label{app:eq:QPCsme1a}
\end{align}
where 
\begin{subequations}\begin{align}
\hat{H}^{\prime}_\mathrm{qb}(t)&= 
\hat{H}_\mathrm{qb} + {\hbar}\sig_{z} \st{\Tcal(t)}\st{\Xcal(t)}\sin(\theta)/2 ,\\ 
\sqrt{\kappa_0(t)}&=\st{\Xcal(t)}\cos(\theta) ,\\
\sqrt{\kappa_1(t)}&=\st{\Xcal(t)}\sin(\theta) ,\\
\Gamma_{\rmd}(t) &= 
\frac{\kappa_0(t)+\kappa_1(t)}{2}=\frac{\st{\Xcal(t)}^2}{2} .
\end{align} \end{subequations}
Here $\theta = \arg(\mathcal{T}^*\Xcal)$ 
is the relative phase between $\Tcal(t)$ and $\Xcal(t)$.  
The superoperator ${\cal H}[\hat{c}]$ in \Eq{app:eq:QPCsme1a} 
is defined by its action on $\rho$ in \Eq{eq:Hnonlinear}.
%has the following action on $\rho$:
%\begin{equation}
%{\cal H}[\hat A]\rho = \hat{c} \rho +\rho \hat{c}\dg 
%- \an{\hat{c}+\hat{c}\dg}\rho .
%\end{equation}

In \Ref{GoaMilWisSunPRB01} the current through the QPC is 
\begin{equation}
I(t)  = e \cu{
|\Tcal|^2+2\mathrm{Re}[\Tcal^*\Xcal] \an{\hat{n}} 
+  |\Tcal| \xi(t) } ,
\end{equation}
which in our parameterization is (now with time-dependent 
$\Tcal$ and $\Xcal$)
\begin{eqnarray}
I(t)  = e  \st{\Tcal(t)}\sq{\st{\Tcal(t)} + \sqrt{\kappa_0(t)} \ro{1+\an{\sig_{z}}} +  \xi(t) } .
\end{eqnarray}
Here we see that this current comprises two parts: a quantum signal, $J(t)$, that depends on the state of the qubit and 
the noise $\xi(t)$; and a large deterministic classical signal, $I_\mathrm{clas}(t)$.  
The quantum and classical signals are 
\begin{eqnarray}\label{eq:quantumSig}
J(t) &=&  \sqrt{\kappa_0(t)}\an{\sig_{z}}+ \xi(t), \\
I_\mathrm{clas}(t) &=&  e  \st{\Tcal(t)} 
\sq{\sqrt{\kappa_0(t)}+  \st{\Tcal(t)}} ,
\end{eqnarray} 
which allows us to write $I(t)$ as
\begin{eqnarray}
I(t) =  e  \st{\Tcal(t)}   J(t)  + I_\mathrm{clas}(t).
\end{eqnarray}
Using the quantum signal we can rewrite the stochastic master 
equation 
[\Eq{app:eq:QPCsme1a}] as 
\begin{align}
  d\rhoc(t) &= 
  - {\frac{i}{\hbar}}
  \sq{\hat{H}^{\prime}_\mathrm{qb}(t) + \Ham_\mathrm{J}(t),\rhoc(t)} dt
\nn\\
&\phantom{=}
  + \half \Gamma_{\rmd}(t)\D{\sig_{z}}\rhoc(t) dt
  \nlalign{
      %\mathcal{L}\rhoc(t) dt
      + [J(t) -  \sqrt{\kappa_0(t)}\an{\sig_{z}}]dt
	  {  
%\frac{\mathrm{Re}\ro{\mathcal{T}^{*}\mathcal{X}}}{2\st{\mathcal{T}}}
	        \h{\sqrt{\kappa_{0}(t)}\sig_{z}/2}
%\frac{\mathrm{Im}\ro{\mathcal{T}^{*}\mathcal{X}}}{2\st{\mathcal{T}}}
	       }\rhoc(t)},
		   \label{app:eq:QPCsme1b}
	 \end{align}
where 
\begin{align} 
\Ham_\mathrm{J}(t) = -{\hbar}\sig_z
\sq{J(t) -\sqrt{\kappa_{0}(t)}\an{\sig_{z}}}
\sqrt{\kappa_{1}(t)} /2.
\end{align}
Equation (\ref{app:eq:QPCsme1b}) is the normalized 
quantum trajectory an observer would use to describe 
their state of knowledge of the qubit if they had access 
to the quantum signal $J(t)$.  

To derive the {\textit realistic} quantum trajectory, we use 
linear quantum trajectory theory,%
\cite{GoeGraPRA94,WisQSO96,GamWisPRA01,GamWisJOBQSO05} 
so we now recast the above in linear form.  The linear 
quantum trajectory is derived in essentially the same way 
as \Eq{app:eq:QPCsme1} was done in \Ref{GoaMilWisSunPRB01}, 
except that one must use an ostensible distribution for $J$ 
rather then the real signal [Eq. \ref{eq:quantumSig}].  
That is, the possible results $Jdt$ at time $t$ are chosen 
from a Gaussian distribution with mean $\mu$ and variance 
$dt$.  Here $\mu$ is an arbitrary ostensible parameter, 
with the only constraint that the ostensible distribution 
for $J$ must be non-zero when the true distribution is 
non-zero.  
We don't go through the calculation in detail, but the final 
result is equivalent to simply replacing 
$\sqrt{\kappa_0(t)}\an{\sig_{z}}$ with $\mu$ in 
\Eq{app:eq:QPCsme1b}.  That is, the linear quantum trajectory 
is
\begin{align}
   d\rhocbar(t) &= 
   - {\frac{i}{\hbar}}
   \sq{\hat{H}^{\prime}_\mathrm{qb}+ \hat{\bar{H}}_\mathrm{J}(t),\rhocbar(t)} dt
\nn\\
&\phantom{=}
   + \half \Gamma_{\rmd}(t)\D{\sig_{z}}\rhocbar(t) dt
   \nlalign{
       %\mathcal{L}\rhoc(t) dt
	       + [J(t) - \mu]dt
	       {  %\frac{\mathrm{Re}\ro{\mathcal{T}^{*}\mathcal{X}}}{2\st{\mathcal{T}}}
	             \hlin{\sqrt{\kappa_{0}(t)}\sig_{z}/2}
	           %\frac{\mathrm{Im}\ro{\mathcal{T}^{*}\mathcal{X}}}{2\st{\mathcal{T}}}
	            }\rhocbar(t)},
	      \label{app:eq:QPCsme1c}
	\end{align}
where the linear measurement superoperator is 
\begin{eqnarray}
\bar{\cal H}[\hat{c}]\rho = \hat{c}\rho +\rho \hat{c}\dg - 
\mu\rho ,
\end{eqnarray}  
and 
\begin{eqnarray}
\hat{\bar{H}}_\mathrm{J}(t) = -{\hbar}\sig_{z}
\sq{J(t) -\mu}\sqrt{\kappa_{1}(t)} /2.
\end{eqnarray} 
Equation (\ref{app:eq:QPCsme1c}) is \Eq{eq:sme}.

%************************************************************
\section{Derivation of the realistic quantum trajectory 
equation \label{app:RQTderivation}}
In this appendix we derive a realistic quantum trajectory 
equation for the rf-QPC-monitored charge qubit.  The 
derivation closely follows the derivation presented 
in \Ref{OxtetalPRB05} for the conventional QPC.  
As discussed, we ignore the 
phase quadrature of the classical oscillator since it reveals 
no qubit information [see Equations (\ref{eqs:dxdytilde:2})].  
Therefore, we describe the imperfect knowledge of the oscillator 
by a probability distribution for the amplitude quadrature, 
$P(\tilde{x})$.

\subsection{Stochastic Fokker-Planck equation for the oscillator
\label{sec:sfpe}}
For notational simplicity, we express the Langevin equation 
for $\tilde{x}(t)$, \Eq{eq:dxtilde:2} (after the deterministic 
classical part has been removed), as
\begin{align}
\rmd \tilde{x}(t) 
&= \sq{m_{\tilde{x}}
+ B_\mathrm{sin}{J}^{\sin}(t)}\dt,
\label{eq:dxtilde:3}
\end{align}
where we have defined 
\begin{align}
m_{\tilde{x}} &\equiv -\frac{\gamma}{2}\tilde{x}(t).
\label{eq:mxtilde}
\end{align}
Equation (\ref{eq:dxtilde:3}) describes the evolution 
of $\tilde{x}(t)$ for perfect knowledge of ${J}^{\sin}(t)$.  
A realistic observer will not have direct access to the 
idealized quadrature current in \Eq{eq:dxtilde:3}, 
so we find an equation for $P(\tilde{x})$ (see 
\Refs{WarWisJOBQSO03a} and \onlinecite{OxtetalPRB05} 
for details of the procedure).  
The result is the stochastic Fokker-Planck equation:
\begin{align}
\rmd {P}_\mathrm{c}(\tilde{x}) &=
\rmd t\left\{-\frac{\du}{\du \tilde{x}}
            \sq{m_{\tilde{x}}
              +B_\mathrm{sin}{J}^{\sin}(t)
               }
      \right.
\nn\\ &\phantom{=+\dt}
      \left.
          +\frac{B^2_\mathrm{sin}}{2}
           \frac{\du^{2}}{\du\tilde{x}^{2}} 
      \right\}P(\tilde{x}) ,
\label{eq:sfpe}
\end{align} 
%{\red What u had
%{\pur\sc\small Jay: I think that this should this be 
%$\rmd\bar{P}$ (unnormalized), since we are using 
%${\tilde{I}}^{\sin}$, not pure white noise.  
%What do you think?\\
%On second thought, its probably correct as it is, since it 
%should just add more deterministic drift.  I'm not sure, 
%so what do you reckon?}\\
%where $S^\mathrm{{\tilde{I}}} = 
%({\tilde{I}}^{\sin}\dt)^{2}/\dt$ 
%%= e^{2}\st{\Tcal_{0}}^{2}S_\mathrm{in}^\mathrm{J} 
%%+ S^\mathrm{{\tilde{I}}\sin}_\mathrm{clas}
%%with $S_\mathrm{in}^\mathrm{J} = S_\mathrm{obs}
%%=1/2 + \epsilon_\mathrm{in}^{2}/16 + \epsilon_\mathrm{in}^{2}/32
%%\approx 1/2$. 
%is the spectral density of the QPC quadrature current 
%${\tilde{I}}^{\sin}$.}
As expected, we have both deterministic and stochastic drift 
($\du/\du\tilde{x}$) of $P(\tilde{x})$, as well as the 
diffusion ($\du^{2}/\du\tilde{x}^{2}$) associated with the 
stochastic drift.  
%Equation (\ref{eq:sfpe}) describes the evolution of 
%$P(\tilde{x})$ conditioned on all the processes in 
%the Langevin equation for $\tilde{x}(t)$, \Eq{eq:dxtilde:3}.  
%%At this stage, 
%%\Eq{eq:sfpe} preserves a delta function initial condition.  
%%That is, for perfect initial knowledge, Equations 
%%(\ref{eq:sfpe}) and (\ref{eq:dxtilde:3}) are equivalent 
%%expressions of the stochastic dynamics of $\tilde{x}(t)$ 
%%for the unrealistic situation of perfect knowledge of the 
%%microscopic processes occurring within the measurement 
%%circuit.  
We next consider conditioning $P(\tilde{x})$ 
on the homodyne measurement results.

\subsection{Zakai equation for the oscillator
\label{sec:zakai}}
Following \Ref{OxtetalPRB05}, we find the best estimate 
for $P(\tilde{x})$ conditioned upon the measurement result 
$\V$ using Bayesian analysis.  
Denoted $P_{\V}(\tilde{x})$, this estimate is
\begin{equation}
\bar{P}_{\V{}}(\tilde{x}) = 
\frac{P_{\tilde{x}}(\V{})P(\tilde{x})}
     {\Lambda (\V{}) } ,
\label{eq:bayes1:rf}
\end{equation}
where the bar indicates an unnormalized distribution.  
The ostensible distribution $\Lambda(\V{})$ is 
a Gaussian distribution of arbitrary mean $\lambda$, and 
some variance $\var$:
\begin{equation}
\Lambda(\V{}) = 
\frac{1}{\sqrt{2\pi\var}}
\exp\sq{-\frac{\ro{\V{}-\lambda}^{2}}
              {2\var}} .
\label{eq:LambdaVx}
\end{equation}
That is, for the Zakai equation we consider the observed 
output $\V{}$ to be ostensibly Gaussian white 
noise of mean $\lambda$.  We simplify the derivation by 
choosing $\lambda=0$, but note that the choice of 
ostensible mean (in $\Lambda$) in the Zakai equation is 
arbitrary {(subject to the condition that $\Lambda(\V)$ is 
nonzero when $P(\V)$ is nonzero)}.  
Inspection of \Eq{eq:Vcal} shows that 
$P_{\tilde{x}}(\V{})$ is a Gaussian 
distribution of mean $\lambda_{\tilde{x}}$ and variance 
$\var = 1/\dt $.  That is,
\begin{align}
P_{\tilde{x}}(\V{}) &=
\frac{\sqrt{\dt}}{\sqrt{2\pi}}
\exp\cu{-\frac{\sq{\V{}-\lambda_{\tilde{x}}}^{2}\dt}
              {2}} ,
\label{eq:Px}
\end{align}
{ where $\lambda_{\tilde{x}}$ is defined in \Eq{eq:lambda}.}
The results above combine into \Eq{eq:bayes1:rf} to 
give the Zakai equation
\begin{align}
\bar{P}_{\V{}}(\tilde{x}) 
&=
\cu{1 + \V{}\rmd t 
        \lambda_{\tilde{x}}
   }
P(\tilde{x}) .
\label{eq:ze}
\end{align} 
This is exactly analogous to the Zakai equations derived 
for the conventional (dc-mode) QPC in \Ref{OxtetalPRB05}.

\subsection{Combined equation for the oscillator
\label{sec:combine}}
As in \Ref{OxtetalPRB05}, we choose to condition on 
measurement results after microscopic processes, so that 
the combined conditional evolution of the oscillator is 
given by
\begin{equation}
{P}(\tilde{x})
 + \rmd \bar{P}_{\V{},\mathrm{c}}(\tilde{x}) = 
\cu{1 + \V{}\rmd t \lambda_{\tilde{x}}
   }
\sq{P(\tilde{x}) + \rmd P_\mathrm{c}(\tilde{x})
   } ,
\label{eq:combine}
\end{equation}
where $\rmd P_\mathrm{c}(\tilde{x})$ is given by 
the stochastic Fokker-Planck equation (\ref{eq:sfpe}).  
We find that
\begin{align}
\rmd \bar{P}_{\V{},\mathrm{c}}(\tilde{x}) 
&= 
\dt\bigg\{
  -\frac{\du}{\du\tilde{x}}
   \sq{m_{\tilde{x}}
      +B_\mathrm{sin}{J}^\mathrm{\sin}(t)}
  +\frac{B_\mathrm{sin}^{2}}{2}
    \frac{\du^{2}}{\du\tilde{x}^{2}}
\bigg.
\nn\\
&\phantom{==}
\bigg.
  +\V{}  \lambda_{\tilde{x}}
\bigg\}P(\tilde{x}) . 
\label{eq:combined:homo}
\end{align}
This equation describes the evolution of the oscillator 
state (via the amplitude quadrature) conditioned on both 
observed and unobserved processes.

\subsection{Stochastic equation for the joint 
qubit-oscillator state
\label{sec:joint}}
The joint system (supersystem) evolution conditioned by 
all processes is found by joining the stochastic master 
equation [\Eq{eq:QPCsmeRWA:normed}] and the combined 
classical evolution [\Eq{eq:combined:homo}].  
This step is performed via\cite{WarWisJOBQSO03a,OxtetalPRB05} 
\begin{align}
\bar{\rho}_{\V{},\mathrm{c}}(\tilde{x})
+\rmd \bar{\rho}_{\V{},\mathrm{c}}(\tilde{x}) 
&= 
\sq{\bar{P}_{\V{},\mathrm{c}}(\tilde{x})
   +\rmd\bar{P}_{\V{},\mathrm{c}}(\tilde{x})}
\nn\\
&\phantom{=}
\times
\sq{\bar{\rho}_\mathrm{c}(t) 
+ \rmd \bar{\rho}_\mathrm{c}(t)} .
\label{eq:joint1}
\end{align}

The coupled noise process in 
$\rmd \bar{P}_{\V{},\mathrm{c}}$ [\Eq{eq:combined:homo}] 
and 
$\rmd \bar{\rho}_\mathrm{c}$ 
[\Eq{eq:QPCsmeRWA:linear}] 
is $J^{\sin} (t)$.  
This is the term from which correlations between 
the circuit and qubit evolution arise in our 
joint stochastic equation.  
This equation describes the evolution of the joint 
quantum-classical state conditioned on the observed 
process $\V(t)$ and the unobserved process 
$J^{\sin} (t)$.  To remove the conditioning 
on the unobserved process, we simply average over it.  
In our approach, the unobserved process is ostensibly 
a white noise of zero mean (since we use the linear 
equation).  
The result is the following superoperator Zakai equation 
\begin{align}
\rmd \bar{\rho}_{\V{}}(\tilde{x}) 
&= 
\dt
\Bigg\{
  -\frac{\du}{\du\tilde{x}}
        m_{\tilde{x}}
  +\frac{B_\mathrm{sin}^{2}}{2}
    \frac{\du^{2}}{\du\tilde{x}^{2}}
  +\V{}\lambda_{\tilde{x}}
+\tilde{\mathcal{L}}
\Bigg\} 
\bar{\rho}_{\V{}}(\tilde{x})
\nn\\
&\phantom{==}
  -\dt
   \sqrt{\eta\kappa}B_{\mathrm{sin}}
%   \ortega_\mathrm{y}
%   \epsilon_\mathrm{in}
%   e\st{\Tcal_{0}}
%   S^\mathrm{J}_{\sin}
  \frac{\du}{\du\tilde{x}}\left[\sig_{z}\bar{\rho}_{\V{}}(\tilde{x})+
\bar{\rho}_{\V{}}(\tilde{x})\sig_{z}\right]/2 ,
\label{eq:superzakai}
\end{align} 
where $\eta$ is defined in \Eq{eq:eta}, 
and $\tilde{\mathcal{L}}$ is defined in \Eq{eq:Ltilde}.

This description of the joint quantum-classical system 
evolution conditioned on the observed output $\V{}(t)$ 
does not preserve normalization of the state.  This is 
because %, in \Sec{sec:zakai}, 
ostensible statistics were chosen for $\V{}(t)$.  
To complete the realistic quantum trajectory derivation, 
we must next find the true statistics for $\V{}(t)$, 
and normalize the superoperator Zakai equation.

%**************************************************************** 
\subsection{Superoperator Kushner-Stratonovich equation
\label{sec:norm}}
Normalization of the superoperator Zakai equation is 
performed\cite{OxtetalPRB05} by taking the trace over 
the qubit and integrating over the oscillator 
($\tilde{x}$):
\begin{equation}
\rho_{\V{}}(\tilde{x})
+\rmd \rho_{\V{}}(\tilde{x}) 
= 
\frac{\bar{\rho}_{\V{}}(\tilde{x})
+\rmd \bar{\rho}_{\V{}}(\tilde{x})}
{\int \tr{\bar{\rho}_{\V{}}(\tilde{x})
+\rmd \bar{\rho}_{\V{}}(\tilde{x})}
d\tilde{x}} .
\label{eq:normalize}
\end{equation}
It is important to note that the resulting equation is not 
the superoperator Kushner-Stratonovich (SKS) equation 
because ostensible statistics are still being used for 
the realistic homodyne output $\V{}(t)$.  
To find the SKS equation, we must substitute the true 
expression for $\V{}\rmd t$.  
The true distribution for $\V{}$ is found from the 
superoperator Zakai equation (\ref{eq:superzakai}) 
by tracing over the qubit and integrating over all 
$\tilde{x}$, then multiplying the result by the ostensible 
distribution $\Lambda(\V{})$ [\Eq{eq:LambdaVx}].  
That is,
\begin{align}
P(\V{}) &= 
\Lambda(\V{})
\int \tr{\bar{\rho}_{\V{}}(\tilde{x})
         + \rmd \bar{\rho}_{\V{}}(\tilde{x})}d\tilde{x}
\nn\\
&=
\frac{\sqrt{\dt}}{\sqrt{2\pi}}
\exp\cu{-{\sq{\V{}-\lambda_{\an{\tilde{x}}_{\rho}}}^{2}\dt}
              /{2}
        } ,
\label{eq:truePV:homo}
\end{align}
{ where 
$\lambda_{\an{\tilde{x}}_{\rho}} \equiv 
\sqrt{A_\mathrm{SN}}
\gamma\an{\tilde{x}}_\rho/ 2B_\mathrm{sin}
$ [c.f. \Eq{eq:lambda}], and 
the average 
$\an{\tilde{x}}_{\rho}
 =\int\tilde{x}\tr{\rho(\tilde{x})}\rmd \tilde{x}$ is 
qubit-dependent (hence the $\rho$ subscript).}  
This means that the true expression for $\V{}(t)$ is
\begin{equation}
\V{}(t)\rmd t = 
\lambda_{\an{\tilde{x}}_{\rho}} \rmd t 
+ \rmd {\cal W}(t) ,
\label{eq:observed:homo:true}
\end{equation}
where $\rmd {\cal W}(t)$ is the white noise 
Wiener increment a realistic observer would see.  The end result is the SKS equation 
(\ref{eq:skse:rfqpc}).

%%%%%%%%%%%%%%%%%%%%%%%%%%%%%%%%%%%%%%%%%%%%%%%%%%%%%%%%%%%%%%%%
\bibliography{rfqpc}

\begin{thebibliography}{61}
\expandafter\ifx\csname natexlab\endcsname\relax\def\natexlab#1{#1}\fi
\expandafter\ifx\csname bibnamefont\endcsname\relax
  \def\bibnamefont#1{#1}\fi
\expandafter\ifx\csname bibfnamefont\endcsname\relax
  \def\bibfnamefont#1{#1}\fi
\expandafter\ifx\csname citenamefont\endcsname\relax
  \def\citenamefont#1{#1}\fi
\expandafter\ifx\csname url\endcsname\relax
  \def\url#1{\texttt{#1}}\fi
\expandafter\ifx\csname urlprefix\endcsname\relax\def\urlprefix{URL }\fi
\providecommand{\bibinfo}[2]{#2}
\providecommand{\eprint}[2][]{\url{#2}}

\bibitem[{\citenamefont{Kane}(1998)}]{KanNAT98}
\bibinfo{author}{\bibfnamefont{B.~E.} \bibnamefont{Kane}},
  \bibinfo{journal}{Nature\ (London)} \textbf{\bibinfo{volume}{393}},
  \bibinfo{pages}{133} (\bibinfo{year}{1998}).

\bibitem[{\citenamefont{Loss and DiVincenzo}(1998)}]{LosDiVPRA98}
\bibinfo{author}{\bibfnamefont{D.}~\bibnamefont{Loss}} \bibnamefont{and}
  \bibinfo{author}{\bibfnamefont{D.~P.} \bibnamefont{DiVincenzo}},
  \bibinfo{journal}{Phys.\ Rev.\ A} \textbf{\bibinfo{volume}{57}},
  \bibinfo{pages}{120} (\bibinfo{year}{1998}).

\bibitem[{\citenamefont{Privman et~al.}(1998)\citenamefont{Privman, Vagner, and
  Kventsel}}]{PriVagKvePLA98}
\bibinfo{author}{\bibfnamefont{V.}~\bibnamefont{Privman}},
  \bibinfo{author}{\bibfnamefont{I.~D.} \bibnamefont{Vagner}},
  \bibnamefont{and} \bibinfo{author}{\bibfnamefont{G.}~\bibnamefont{Kventsel}},
  \bibinfo{journal}{Phys.\ Lett.\ A} \textbf{\bibinfo{volume}{239}},
  \bibinfo{pages}{141} (\bibinfo{year}{1998}).

\bibitem[{\citenamefont{Imamo{\=g}lu et~al.}(1999)\citenamefont{Imamo{\=g}lu,
  Awschalom, Burkard, DiVincenzo, Loss, Sherwin, and Small}}]{ImaetalPRL99}
\bibinfo{author}{\bibfnamefont{A.}~\bibnamefont{Imamo{\=g}lu}},
  \bibinfo{author}{\bibfnamefont{D.~D.} \bibnamefont{Awschalom}},
  \bibinfo{author}{\bibfnamefont{G.}~\bibnamefont{Burkard}},
  \bibinfo{author}{\bibfnamefont{D.~P.} \bibnamefont{DiVincenzo}},
  \bibinfo{author}{\bibfnamefont{D.}~\bibnamefont{Loss}},
  \bibinfo{author}{\bibfnamefont{M.}~\bibnamefont{Sherwin}}, \bibnamefont{and}
  \bibinfo{author}{\bibfnamefont{A.}~\bibnamefont{Small}},
  \bibinfo{journal}{Phys.\ Rev.\ Lett.} \textbf{\bibinfo{volume}{83}},
  \bibinfo{pages}{4204} (\bibinfo{year}{1999}).

\bibitem[{\citenamefont{Vrijen et~al.}(2000)\citenamefont{Vrijen, Yablonovitch,
  Wang, Jiang, Balandin, Roychowdhury, Mor, and DiVincenzo}}]{VrietalPRA00}
\bibinfo{author}{\bibfnamefont{R.}~\bibnamefont{Vrijen}},
  \bibinfo{author}{\bibfnamefont{E.}~\bibnamefont{Yablonovitch}},
  \bibinfo{author}{\bibfnamefont{K.}~\bibnamefont{Wang}},
  \bibinfo{author}{\bibfnamefont{H.~W.} \bibnamefont{Jiang}},
  \bibinfo{author}{\bibfnamefont{A.}~\bibnamefont{Balandin}},
  \bibinfo{author}{\bibfnamefont{V.}~\bibnamefont{Roychowdhury}},
  \bibinfo{author}{\bibfnamefont{T.}~\bibnamefont{Mor}}, \bibnamefont{and}
  \bibinfo{author}{\bibfnamefont{D.}~\bibnamefont{DiVincenzo}},
  \bibinfo{journal}{Phys.\ Rev.\ A} \textbf{\bibinfo{volume}{62}},
  \bibinfo{pages}{012306} (\bibinfo{year}{2000}).

\bibitem[{\citenamefont{Carmichael}(1993)}]{OpenSys}
\bibinfo{author}{\bibfnamefont{H.~J.} \bibnamefont{Carmichael}},
  \emph{\bibinfo{title}{An Open Systems Approach to Quantum Optics}}
  (\bibinfo{publisher}{Springer}, \bibinfo{address}{Berlin},
  \bibinfo{year}{1993}).

\bibitem[{\citenamefont{Wiseman and
  Milburn}(1993{\natexlab{a}})}]{WisMilPRA93b}
\bibinfo{author}{\bibfnamefont{H.~M.} \bibnamefont{Wiseman}} \bibnamefont{and}
  \bibinfo{author}{\bibfnamefont{G.~J.} \bibnamefont{Milburn}},
  \bibinfo{journal}{Phys.\ Rev.\ A} \textbf{\bibinfo{volume}{47}},
  \bibinfo{pages}{1652} (\bibinfo{year}{1993}{\natexlab{a}}).

\bibitem[{\citenamefont{Wiseman}(1996)}]{WisQSO96}
\bibinfo{author}{\bibfnamefont{H.~M.} \bibnamefont{Wiseman}},
  \bibinfo{journal}{Quantum\ Semiclass.\ Opt.} \textbf{\bibinfo{volume}{8}},
  \bibinfo{pages}{205} (\bibinfo{year}{1996}).

\bibitem[{\citenamefont{Korotkov}(1999)}]{KorPRB99}
\bibinfo{author}{\bibfnamefont{A.~N.} \bibnamefont{Korotkov}},
  \bibinfo{journal}{Phys.\ Rev.\ B} \textbf{\bibinfo{volume}{60}},
  \bibinfo{pages}{5737} (\bibinfo{year}{1999}).

\bibitem[{\citenamefont{Korotkov}(2001{\natexlab{a}})}]{KorPRB01b}
\bibinfo{author}{\bibfnamefont{A.~N.} \bibnamefont{Korotkov}},
  \bibinfo{journal}{Phys.\ Rev.\ B} \textbf{\bibinfo{volume}{63}},
  \bibinfo{pages}{115403} (\bibinfo{year}{2001}{\natexlab{a}}).

\bibitem[{\citenamefont{Wiseman et~al.}(2001)\citenamefont{Wiseman, {Wahyu
  Utami}, Sun, Milburn, Kane, Dzurak, and Clark}}]{WisetalPRB01}
\bibinfo{author}{\bibfnamefont{H.~M.} \bibnamefont{Wiseman}},
  \bibinfo{author}{\bibfnamefont{D.}~\bibnamefont{{Wahyu Utami}}},
  \bibinfo{author}{\bibfnamefont{H.~B.} \bibnamefont{Sun}},
  \bibinfo{author}{\bibfnamefont{G.~J.} \bibnamefont{Milburn}},
  \bibinfo{author}{\bibfnamefont{B.~E.} \bibnamefont{Kane}},
  \bibinfo{author}{\bibfnamefont{A.}~\bibnamefont{Dzurak}}, \bibnamefont{and}
  \bibinfo{author}{\bibfnamefont{R.~G.} \bibnamefont{Clark}},
  \bibinfo{journal}{Phys.\ Rev.\ B} \textbf{\bibinfo{volume}{63}},
  \bibinfo{pages}{235308} (\bibinfo{year}{2001}).

\bibitem[{\citenamefont{Goan et~al.}(2001)\citenamefont{Goan, Milburn, Wiseman,
  and Sun}}]{GoaMilWisSunPRB01}
\bibinfo{author}{\bibfnamefont{H.-S.} \bibnamefont{Goan}},
  \bibinfo{author}{\bibfnamefont{G.~J.} \bibnamefont{Milburn}},
  \bibinfo{author}{\bibfnamefont{H.~M.} \bibnamefont{Wiseman}},
  \bibnamefont{and} \bibinfo{author}{\bibfnamefont{H.~B.} \bibnamefont{Sun}},
  \bibinfo{journal}{Phys.\ Rev.\ B} \textbf{\bibinfo{volume}{63}},
  \bibinfo{pages}{125326} (\bibinfo{year}{2001}).

\bibitem[{\citenamefont{Korotkov}(2001{\natexlab{b}})}]{KorPRB01c}
\bibinfo{author}{\bibfnamefont{A.~N.} \bibnamefont{Korotkov}},
  \bibinfo{journal}{Phys.\ Rev.\ B} \textbf{\bibinfo{volume}{64}},
  \bibinfo{pages}{193407} (\bibinfo{year}{2001}{\natexlab{b}}).

\bibitem[{\citenamefont{Goan and Milburn}(2001)}]{GoaMilPRB01}
\bibinfo{author}{\bibfnamefont{H.-S.} \bibnamefont{Goan}} \bibnamefont{and}
  \bibinfo{author}{\bibfnamefont{G.~J.} \bibnamefont{Milburn}},
  \bibinfo{journal}{Phys.\ Rev.\ B} \textbf{\bibinfo{volume}{64}},
  \bibinfo{pages}{235307} (\bibinfo{year}{2001}).

\bibitem[{\citenamefont{Korotkov}(2003{\natexlab{a}})}]{KorPRB03}
\bibinfo{author}{\bibfnamefont{A.~N.} \bibnamefont{Korotkov}},
  \bibinfo{journal}{Phys.\ Rev.\ B} \textbf{\bibinfo{volume}{67}},
  \bibinfo{pages}{235408} (\bibinfo{year}{2003}{\natexlab{a}}).

\bibitem[{\citenamefont{Korotkov}(2003{\natexlab{b}})}]{Kor03}
\bibinfo{author}{\bibfnamefont{A.~N.} \bibnamefont{Korotkov}}, in
  \emph{\bibinfo{booktitle}{Quantum Noise in Mesoscopic Physics}}, edited by
  \bibinfo{editor}{\bibfnamefont{Y.~V.} \bibnamefont{Nazarov}}
  (\bibinfo{publisher}{Kluwer Academic}, \bibinfo{address}{Netherlands},
  \bibinfo{year}{2003}{\natexlab{b}}), pp. \bibinfo{pages}{205--228}.

\bibitem[{\citenamefont{Stace and Barrett}(2004{\natexlab{a}})}]{StaBar-04}
\bibinfo{author}{\bibfnamefont{T.~M.} \bibnamefont{Stace}} \bibnamefont{and}
  \bibinfo{author}{\bibfnamefont{S.~D.} \bibnamefont{Barrett}}
  (\bibinfo{year}{2004}{\natexlab{a}}), \bibinfo{note}{cond-mat/0309610}.

\bibitem[{\citenamefont{Stace and Barrett}(2004{\natexlab{b}})}]{StaBarPRL04}
\bibinfo{author}{\bibfnamefont{T.~M.} \bibnamefont{Stace}} \bibnamefont{and}
  \bibinfo{author}{\bibfnamefont{S.~D.} \bibnamefont{Barrett}},
  \bibinfo{journal}{Phys.\ Rev.\ Lett.} \textbf{\bibinfo{volume}{92}},
  \bibinfo{pages}{136802} (\bibinfo{year}{2004}{\natexlab{b}}).

\bibitem[{\citenamefont{Goan}(2004)}]{GoaPRB04}
\bibinfo{author}{\bibfnamefont{H.-S.} \bibnamefont{Goan}},
  \bibinfo{journal}{Phys.\ Rev.\ B} \textbf{\bibinfo{volume}{70}},
  \bibinfo{pages}{075305} (\bibinfo{year}{2004}).

\bibitem[{\citenamefont{Gambetta et~al.}(2008)\citenamefont{Gambetta, Blais,
  Boissonneault, Houck, Schuster, and Girvin}}]{GametalPRA08}
\bibinfo{author}{\bibfnamefont{J.}~\bibnamefont{Gambetta}},
  \bibinfo{author}{\bibfnamefont{A.}~\bibnamefont{Blais}},
  \bibinfo{author}{\bibfnamefont{M.}~\bibnamefont{Boissonneault}},
  \bibinfo{author}{\bibfnamefont{A.~A.} \bibnamefont{Houck}},
  \bibinfo{author}{\bibfnamefont{D.~I.} \bibnamefont{Schuster}},
  \bibnamefont{and} \bibinfo{author}{\bibfnamefont{S.~M.}
  \bibnamefont{Girvin}}, \bibinfo{journal}{Physical Review A}
  \textbf{\bibinfo{volume}{77}}, \bibinfo{eid}{012112} (\bibinfo{year}{2008}).

\bibitem[{\citenamefont{Oxtoby et~al.}(2005)\citenamefont{Oxtoby, Warszawski,
  Wiseman, Sun, and Polkinghorne}}]{OxtetalPRB05}
\bibinfo{author}{\bibfnamefont{N.~P.} \bibnamefont{Oxtoby}},
  \bibinfo{author}{\bibfnamefont{P.}~\bibnamefont{Warszawski}},
  \bibinfo{author}{\bibfnamefont{H.~M.} \bibnamefont{Wiseman}},
  \bibinfo{author}{\bibfnamefont{H.-B.} \bibnamefont{Sun}}, \bibnamefont{and}
  \bibinfo{author}{\bibfnamefont{R.~E.~S.} \bibnamefont{Polkinghorne}},
  \bibinfo{journal}{Phys.\ Rev.\ B} \textbf{\bibinfo{volume}{71}},
  \bibinfo{pages}{165317} (\bibinfo{year}{2005}).

\bibitem[{\citenamefont{Warszawski et~al.}(2002)\citenamefont{Warszawski,
  Wiseman, and Mabuchi}}]{WarWisMabPRA02}
\bibinfo{author}{\bibfnamefont{P.}~\bibnamefont{Warszawski}},
  \bibinfo{author}{\bibfnamefont{H.~M.} \bibnamefont{Wiseman}},
  \bibnamefont{and} \bibinfo{author}{\bibfnamefont{H.}~\bibnamefont{Mabuchi}},
  \bibinfo{journal}{Phys.\ Rev.\ A} \textbf{\bibinfo{volume}{65}},
  \bibinfo{pages}{023802} (\bibinfo{year}{2002}).

\bibitem[{\citenamefont{Warszawski and Wiseman}(2003)}]{WarWisJOBQSO03a}
\bibinfo{author}{\bibfnamefont{P.}~\bibnamefont{Warszawski}} \bibnamefont{and}
  \bibinfo{author}{\bibfnamefont{H.~M.} \bibnamefont{Wiseman}},
  \bibinfo{journal}{J.\ Opt.\ B:\ Quantum\ Semiclass.\ Opt.}
  \textbf{\bibinfo{volume}{5}}, \bibinfo{pages}{1} (\bibinfo{year}{2003}).

\bibitem[{\citenamefont{Zimmerli et~al.}(1992)\citenamefont{Zimmerli, Eiles,
  Kautz, and Martinis}}]{ZimetalAPL92}
\bibinfo{author}{\bibfnamefont{G.}~\bibnamefont{Zimmerli}},
  \bibinfo{author}{\bibfnamefont{T.~M.} \bibnamefont{Eiles}},
  \bibinfo{author}{\bibfnamefont{R.~L.} \bibnamefont{Kautz}}, \bibnamefont{and}
  \bibinfo{author}{\bibfnamefont{J.~M.} \bibnamefont{Martinis}},
  \bibinfo{journal}{Appl.\ Phys.\ Lett.} \textbf{\bibinfo{volume}{61}},
  \bibinfo{pages}{237} (\bibinfo{year}{1992}).

\bibitem[{\citenamefont{Wong}(2003)}]{WonMR03}
\bibinfo{author}{\bibfnamefont{H.}~\bibnamefont{Wong}},
  \bibinfo{journal}{Microelectronics Reliability}
  \textbf{\bibinfo{volume}{43}}, \bibinfo{pages}{585} (\bibinfo{year}{2003}).

\bibitem[{\citenamefont{Schoelkopf et~al.}(1998)\citenamefont{Schoelkopf,
  Wahlgren, Kozhevnikov, Delsing, and Prober}}]{SchetalSC98}
\bibinfo{author}{\bibfnamefont{R.~J.} \bibnamefont{Schoelkopf}},
  \bibinfo{author}{\bibfnamefont{P.}~\bibnamefont{Wahlgren}},
  \bibinfo{author}{\bibfnamefont{A.~A.} \bibnamefont{Kozhevnikov}},
  \bibinfo{author}{\bibfnamefont{P.}~\bibnamefont{Delsing}}, \bibnamefont{and}
  \bibinfo{author}{\bibfnamefont{D.~E.} \bibnamefont{Prober}},
  \bibinfo{journal}{Science} \textbf{\bibinfo{volume}{280}},
  \bibinfo{pages}{1238} (\bibinfo{year}{1998}).

\bibitem[{\citenamefont{Aassime
  et~al.}(2001{\natexlab{a}})\citenamefont{Aassime, Johansson, Wendin,
  Schoelkopf, and Delsing}}]{AasetalPRL01}
\bibinfo{author}{\bibfnamefont{A.}~\bibnamefont{Aassime}},
  \bibinfo{author}{\bibfnamefont{G.}~\bibnamefont{Johansson}},
  \bibinfo{author}{\bibfnamefont{G.}~\bibnamefont{Wendin}},
  \bibinfo{author}{\bibfnamefont{R.~J.} \bibnamefont{Schoelkopf}},
  \bibnamefont{and} \bibinfo{author}{\bibfnamefont{P.}~\bibnamefont{Delsing}},
  \bibinfo{journal}{Phys.\ Rev.\ Lett.} \textbf{\bibinfo{volume}{86}},
  \bibinfo{pages}{3376} (\bibinfo{year}{2001}{\natexlab{a}}).

\bibitem[{\citenamefont{Qin and Williams}(2006)}]{QinWilAPL06}
\bibinfo{author}{\bibfnamefont{H.}~\bibnamefont{Qin}} \bibnamefont{and}
  \bibinfo{author}{\bibfnamefont{D.~A.} \bibnamefont{Williams}},
  \bibinfo{journal}{Appl.\ Phys.\ Lett.} \textbf{\bibinfo{volume}{88}},
  \bibinfo{pages}{203506} (\bibinfo{year}{2006}).

\bibitem[{\citenamefont{Gambetta and Wiseman}(2005)}]{GamWisJOBQSO05}
\bibinfo{author}{\bibfnamefont{J.}~\bibnamefont{Gambetta}} \bibnamefont{and}
  \bibinfo{author}{\bibfnamefont{H.~M.} \bibnamefont{Wiseman}},
  \bibinfo{journal}{J.\ Opt.\ B:\ Quantum\ Semiclass.\ Opt.}
  \textbf{\bibinfo{volume}{7}}, \bibinfo{pages}{S250} (\bibinfo{year}{2005}).

\bibitem[{\citenamefont{Ahn et~al.}(2003)\citenamefont{Ahn, Wiseman, and
  Milburn}}]{AhnWisMilPRA03}
\bibinfo{author}{\bibfnamefont{C.}~\bibnamefont{Ahn}},
  \bibinfo{author}{\bibfnamefont{H.~M.} \bibnamefont{Wiseman}},
  \bibnamefont{and} \bibinfo{author}{\bibfnamefont{G.~J.}
  \bibnamefont{Milburn}}, \bibinfo{journal}{Phys.\ Rev.\ A}
  \textbf{\bibinfo{volume}{67}}, \bibinfo{pages}{052310}
  (\bibinfo{year}{2003}).

\bibitem[{\citenamefont{Sarovar et~al.}(2004)\citenamefont{Sarovar, Ahn,
  Jacobs, and Milburn}}]{SarAhnJacMilPRA04}
\bibinfo{author}{\bibfnamefont{M.}~\bibnamefont{Sarovar}},
  \bibinfo{author}{\bibfnamefont{C.}~\bibnamefont{Ahn}},
  \bibinfo{author}{\bibfnamefont{K.}~\bibnamefont{Jacobs}}, \bibnamefont{and}
  \bibinfo{author}{\bibfnamefont{G.~J.} \bibnamefont{Milburn}},
  \bibinfo{journal}{Phys.\ Rev.\ A} \textbf{\bibinfo{volume}{69}},
  \bibinfo{pages}{052324} (\bibinfo{year}{2004}).

\bibitem[{\citenamefont{Ahn et~al.}(2004)\citenamefont{Ahn, Wiseman, and
  Jacobs}}]{AhnWisJacPRA04}
\bibinfo{author}{\bibfnamefont{C.}~\bibnamefont{Ahn}},
  \bibinfo{author}{\bibfnamefont{H.~M.} \bibnamefont{Wiseman}},
  \bibnamefont{and} \bibinfo{author}{\bibfnamefont{K.}~\bibnamefont{Jacobs}},
  \bibinfo{journal}{Phys.\ Rev.\ A} \textbf{\bibinfo{volume}{70}},
  \bibinfo{pages}{024302} (\bibinfo{year}{2004}).

\bibitem[{\citenamefont{van Handel and Mabuchi}(2005)}]{vHanMab05}
\bibinfo{author}{\bibfnamefont{R.}~\bibnamefont{van Handel}} \bibnamefont{and}
  \bibinfo{author}{\bibfnamefont{H.}~\bibnamefont{Mabuchi}}
  (\bibinfo{year}{2005}), \bibinfo{note}{quant-ph/0511221}.

\bibitem[{\citenamefont{Wiseman and Milburn}(1993{\natexlab{b}})}]{WisMilPRL93}
\bibinfo{author}{\bibfnamefont{H.~M.} \bibnamefont{Wiseman}} \bibnamefont{and}
  \bibinfo{author}{\bibfnamefont{G.~J.} \bibnamefont{Milburn}},
  \bibinfo{journal}{Phys.\ Rev.\ Lett.} \textbf{\bibinfo{volume}{70}},
  \bibinfo{pages}{548} (\bibinfo{year}{1993}{\natexlab{b}}).

\bibitem[{\citenamefont{Wiseman}(1995)}]{WisPRL95}
\bibinfo{author}{\bibfnamefont{H.~M.} \bibnamefont{Wiseman}},
  \bibinfo{journal}{Phys.\ Rev.\ Lett.} \textbf{\bibinfo{volume}{75}},
  \bibinfo{pages}{4587} (\bibinfo{year}{1995}).

\bibitem[{\citenamefont{Doherty and Jacobs}(1999)}]{DohJacPRA99}
\bibinfo{author}{\bibfnamefont{A.~C.} \bibnamefont{Doherty}} \bibnamefont{and}
  \bibinfo{author}{\bibfnamefont{K.}~\bibnamefont{Jacobs}},
  \bibinfo{journal}{Phys.\ Rev.\ A} \textbf{\bibinfo{volume}{60}},
  \bibinfo{pages}{2700} (\bibinfo{year}{1999}).

\bibitem[{\citenamefont{Doherty et~al.}(2000)\citenamefont{Doherty, Habib,
  Jacobs, Mabuchi, and Tan}}]{DohetalPRA00}
\bibinfo{author}{\bibfnamefont{A.~C.} \bibnamefont{Doherty}},
  \bibinfo{author}{\bibfnamefont{S.}~\bibnamefont{Habib}},
  \bibinfo{author}{\bibfnamefont{K.}~\bibnamefont{Jacobs}},
  \bibinfo{author}{\bibfnamefont{H.}~\bibnamefont{Mabuchi}}, \bibnamefont{and}
  \bibinfo{author}{\bibfnamefont{S.~M.} \bibnamefont{Tan}},
  \bibinfo{journal}{Phys.\ Rev.\ A} \textbf{\bibinfo{volume}{62}},
  \bibinfo{pages}{012105} (\bibinfo{year}{2000}).

\bibitem[{\citenamefont{Armen et~al.}(2002)\citenamefont{Armen, Au, Stockton,
  Doherty, and Mabuchi}}]{ArmetalPRL02}
\bibinfo{author}{\bibfnamefont{M.~A.} \bibnamefont{Armen}},
  \bibinfo{author}{\bibfnamefont{J.~K.} \bibnamefont{Au}},
  \bibinfo{author}{\bibfnamefont{J.~K.} \bibnamefont{Stockton}},
  \bibinfo{author}{\bibfnamefont{A.~C.} \bibnamefont{Doherty}},
  \bibnamefont{and} \bibinfo{author}{\bibfnamefont{H.}~\bibnamefont{Mabuchi}},
  \bibinfo{journal}{Phys.\ Rev.\ Lett.} \textbf{\bibinfo{volume}{89}},
  \bibinfo{pages}{133602} (\bibinfo{year}{2002}).

\bibitem[{\citenamefont{Wiseman et~al.}(2002)\citenamefont{Wiseman, Mancini,
  and Wang}}]{WisManWanPRA02}
\bibinfo{author}{\bibfnamefont{H.~M.} \bibnamefont{Wiseman}},
  \bibinfo{author}{\bibfnamefont{S.}~\bibnamefont{Mancini}}, \bibnamefont{and}
  \bibinfo{author}{\bibfnamefont{J.}~\bibnamefont{Wang}},
  \bibinfo{journal}{Phys.\ Rev.\ A} \textbf{\bibinfo{volume}{66}},
  \bibinfo{pages}{013807} (\bibinfo{year}{2002}).

\bibitem[{\citenamefont{Smith et~al.}(2002)\citenamefont{Smith, Reiner, Orozco,
  Kuhr, and Wiseman}}]{SmietalPRL02}
\bibinfo{author}{\bibfnamefont{W.~P.} \bibnamefont{Smith}},
  \bibinfo{author}{\bibfnamefont{J.~E.} \bibnamefont{Reiner}},
  \bibinfo{author}{\bibfnamefont{L.~A.} \bibnamefont{Orozco}},
  \bibinfo{author}{\bibfnamefont{S.}~\bibnamefont{Kuhr}}, \bibnamefont{and}
  \bibinfo{author}{\bibfnamefont{H.~M.} \bibnamefont{Wiseman}},
  \bibinfo{journal}{Phys.\ Rev.\ Lett.} \textbf{\bibinfo{volume}{89}},
  \bibinfo{pages}{133601} (\bibinfo{year}{2002}).

\bibitem[{\citenamefont{Ruskov and Korotkov}(2002)}]{RusKorPRB02}
\bibinfo{author}{\bibfnamefont{R.}~\bibnamefont{Ruskov}} \bibnamefont{and}
  \bibinfo{author}{\bibfnamefont{A.~N.} \bibnamefont{Korotkov}},
  \bibinfo{journal}{Phys.\ Rev.\ B} \textbf{\bibinfo{volume}{66}},
  \bibinfo{pages}{041401(R)} (\bibinfo{year}{2002}).

\bibitem[{\citenamefont{Jacobs}(2003)}]{JacPRA03}
\bibinfo{author}{\bibfnamefont{K.}~\bibnamefont{Jacobs}},
  \bibinfo{journal}{Phys. Rev. A} \textbf{\bibinfo{volume}{67}},
  \bibinfo{pages}{030301(R)} (\bibinfo{year}{2003}).

\bibitem[{\citenamefont{Combes and Jacobs}(2006)}]{ComJacPRL06}
\bibinfo{author}{\bibfnamefont{J.}~\bibnamefont{Combes}} \bibnamefont{and}
  \bibinfo{author}{\bibfnamefont{K.}~\bibnamefont{Jacobs}},
  \bibinfo{journal}{Phys. Rev. Lett.} \textbf{\bibinfo{volume}{96}},
  \bibinfo{pages}{010504} (\bibinfo{year}{2006}).

\bibitem[{\citenamefont{Aassime
  et~al.}(2001{\natexlab{b}})\citenamefont{Aassime, Gunnarsson, Bladh, Delsing,
  and Schoelkopf}}]{AasetalAPL01}
\bibinfo{author}{\bibfnamefont{A.}~\bibnamefont{Aassime}},
  \bibinfo{author}{\bibfnamefont{D.}~\bibnamefont{Gunnarsson}},
  \bibinfo{author}{\bibfnamefont{K.}~\bibnamefont{Bladh}},
  \bibinfo{author}{\bibfnamefont{P.}~\bibnamefont{Delsing}}, \bibnamefont{and}
  \bibinfo{author}{\bibfnamefont{R.~J.} \bibnamefont{Schoelkopf}},
  \bibinfo{journal}{Appl.\ Phys.\ Lett.} \textbf{\bibinfo{volume}{79}},
  \bibinfo{pages}{4031} (\bibinfo{year}{2001}{\natexlab{b}}).

\bibitem[{\citenamefont{Hayashi et~al.}(2003)\citenamefont{Hayashi, Fujisawa,
  Cheong, Jeong, and Hirayama}}]{HayetalPRL03}
\bibinfo{author}{\bibfnamefont{T.}~\bibnamefont{Hayashi}},
  \bibinfo{author}{\bibfnamefont{T.}~\bibnamefont{Fujisawa}},
  \bibinfo{author}{\bibfnamefont{H.~D.} \bibnamefont{Cheong}},
  \bibinfo{author}{\bibfnamefont{Y.~H.} \bibnamefont{Jeong}}, \bibnamefont{and}
  \bibinfo{author}{\bibfnamefont{Y.}~\bibnamefont{Hirayama}},
  \bibinfo{journal}{Phys.\ Rev.\ Lett.} \textbf{\bibinfo{volume}{91}},
  \bibinfo{pages}{226804} (\bibinfo{year}{2003}).

\bibitem[{\citenamefont{Gorman et~al.}(2005)\citenamefont{Gorman, Hasko, and
  Williams}}]{GorHasWilPRL05}
\bibinfo{author}{\bibfnamefont{J.}~\bibnamefont{Gorman}},
  \bibinfo{author}{\bibfnamefont{D.~G.} \bibnamefont{Hasko}}, \bibnamefont{and}
  \bibinfo{author}{\bibfnamefont{D.~A.} \bibnamefont{Williams}},
  \bibinfo{journal}{Phys.\ Rev.\ Lett.} \textbf{\bibinfo{volume}{95}},
  \bibinfo{pages}{090502} (\bibinfo{year}{2005}).

\bibitem[{\citenamefont{Reilly et~al.}(2007)\citenamefont{Reilly, Marcus,
  Hanson, and Gossard}}]{ReietalAPL07}
\bibinfo{author}{\bibfnamefont{D.~J.} \bibnamefont{Reilly}},
  \bibinfo{author}{\bibfnamefont{C.~M.} \bibnamefont{Marcus}},
  \bibinfo{author}{\bibfnamefont{M.~P.} \bibnamefont{Hanson}},
  \bibnamefont{and} \bibinfo{author}{\bibfnamefont{A.~C.}
  \bibnamefont{Gossard}}, \bibinfo{journal}{Appl.\ Phys.\ Lett.}
  \textbf{\bibinfo{volume}{91}}, \bibinfo{pages}{162101}
  (\bibinfo{year}{2007}).

\bibitem[{\citenamefont{Turin and Korotkov}(2004)}]{TurKorPRB04}
\bibinfo{author}{\bibfnamefont{V.~O.} \bibnamefont{Turin}} \bibnamefont{and}
  \bibinfo{author}{\bibfnamefont{A.~N.} \bibnamefont{Korotkov}},
  \bibinfo{journal}{Phys.\ Rev.\ B} \textbf{\bibinfo{volume}{69}},
  \bibinfo{pages}{195310} (\bibinfo{year}{2004}).

\bibitem[{\citenamefont{Bladh et~al.}(2005)\citenamefont{Bladh, Duty,
  Gunnarsson, and Delsing}}]{BlaetalNJP05}
\bibinfo{author}{\bibfnamefont{K.}~\bibnamefont{Bladh}},
  \bibinfo{author}{\bibfnamefont{T.}~\bibnamefont{Duty}},
  \bibinfo{author}{\bibfnamefont{D.}~\bibnamefont{Gunnarsson}},
  \bibnamefont{and} \bibinfo{author}{\bibfnamefont{P.}~\bibnamefont{Delsing}},
  \bibinfo{journal}{New Journal of Physics} \textbf{\bibinfo{volume}{7}},
  \bibinfo{pages}{180} (\bibinfo{year}{2005}).

\bibitem[{\citenamefont{Wiseman and
  Milburn}(1993{\natexlab{c}})}]{WisMilPRA93a}
\bibinfo{author}{\bibfnamefont{H.~M.} \bibnamefont{Wiseman}} \bibnamefont{and}
  \bibinfo{author}{\bibfnamefont{G.~J.} \bibnamefont{Milburn}},
  \bibinfo{journal}{Phys.\ Rev.\ A} \textbf{\bibinfo{volume}{47}},
  \bibinfo{pages}{642} (\bibinfo{year}{1993}{\natexlab{c}}).

\bibitem[{\citenamefont{Pozar}(1998)}]{Pozar98}
\bibinfo{author}{\bibfnamefont{D.~M.} \bibnamefont{Pozar}},
  \emph{\bibinfo{title}{Microwave Engineering}} (\bibinfo{publisher}{Wiley},
  \bibinfo{address}{New York}, \bibinfo{year}{1998}).

\bibitem[{\citenamefont{Averin and Sukhorukov}(2005)}]{AveSukPRL05}
\bibinfo{author}{\bibfnamefont{D.~V.} \bibnamefont{Averin}} \bibnamefont{and}
  \bibinfo{author}{\bibfnamefont{E.~V.} \bibnamefont{Sukhorukov}},
  \bibinfo{journal}{Phys.\ Rev.\ Lett.} \textbf{\bibinfo{volume}{95}},
  \bibinfo{pages}{126803} (\bibinfo{year}{2005}).

\bibitem[{\citenamefont{Oxtoby et~al.}(2006)\citenamefont{Oxtoby, Wiseman, and
  Sun}}]{OxtWisSunPRB06}
\bibinfo{author}{\bibfnamefont{N.~P.} \bibnamefont{Oxtoby}},
  \bibinfo{author}{\bibfnamefont{H.~M.} \bibnamefont{Wiseman}},
  \bibnamefont{and} \bibinfo{author}{\bibfnamefont{H.-B.} \bibnamefont{Sun}},
  \bibinfo{journal}{Phys.\ Rev.\ B} \textbf{\bibinfo{volume}{74}},
  \bibinfo{pages}{045328} (\bibinfo{year}{2006}).

\bibitem[{\citenamefont{Gardiner and Zoller}(2000)}]{QNoise}
\bibinfo{author}{\bibfnamefont{C.~W.} \bibnamefont{Gardiner}} \bibnamefont{and}
  \bibinfo{author}{\bibfnamefont{P.}~\bibnamefont{Zoller}},
  \emph{\bibinfo{title}{Quantum Noise}} (\bibinfo{publisher}{Springer},
  \bibinfo{address}{Berlin}, \bibinfo{year}{2000}), \bibinfo{edition}{2nd} ed.

\bibitem[{\citenamefont{{Gatarek} and {Gisin}}(1991)}]{GatGisJMP91}
\bibinfo{author}{\bibfnamefont{D.}~\bibnamefont{{Gatarek}}} \bibnamefont{and}
  \bibinfo{author}{\bibfnamefont{N.}~\bibnamefont{{Gisin}}},
  \bibinfo{journal}{Journal of Mathematical Physics}
  \textbf{\bibinfo{volume}{32}}, \bibinfo{pages}{2152} (\bibinfo{year}{1991}).

\bibitem[{\citenamefont{Oxtoby}(2006)}]{OxtPHD07}
\bibinfo{author}{\bibfnamefont{N.~P.} \bibnamefont{Oxtoby}}, Ph.D. thesis,
  \bibinfo{school}{Griffith University} (\bibinfo{year}{2006}).

\bibitem[{\citenamefont{Goetsch and Graham}(1994)}]{GoeGraPRA94}
\bibinfo{author}{\bibfnamefont{P.}~\bibnamefont{Goetsch}} \bibnamefont{and}
  \bibinfo{author}{\bibfnamefont{R.}~\bibnamefont{Graham}},
  \bibinfo{journal}{Phys.\ Rev.\ A} \textbf{\bibinfo{volume}{50}},
  \bibinfo{pages}{5242} (\bibinfo{year}{1994}).

\bibitem[{\citenamefont{Gambetta and Wiseman}(2001)}]{GamWisPRA01}
\bibinfo{author}{\bibfnamefont{J.}~\bibnamefont{Gambetta}} \bibnamefont{and}
  \bibinfo{author}{\bibfnamefont{H.~M.} \bibnamefont{Wiseman}},
  \bibinfo{journal}{Phys.\ Rev.\ A} \textbf{\bibinfo{volume}{64}},
  \bibinfo{pages}{042105} (\bibinfo{year}{2001}).

\bibitem[{\citenamefont{B\"uttiker}(1992)}]{ButPRB92}
\bibinfo{author}{\bibfnamefont{M.}~\bibnamefont{B\"uttiker}},
  \bibinfo{journal}{Phys. Rev. B} \textbf{\bibinfo{volume}{46}},
  \bibinfo{pages}{12485} (\bibinfo{year}{1992}).

\bibitem[{\citenamefont{Fujisawa and Hirayama}(2000)}]{FujHirAPL00}
\bibinfo{author}{\bibfnamefont{T.}~\bibnamefont{Fujisawa}} \bibnamefont{and}
  \bibinfo{author}{\bibfnamefont{Y.}~\bibnamefont{Hirayama}},
  \bibinfo{journal}{Appl.\ Phys.\ Lett.} \textbf{\bibinfo{volume}{77}},
  \bibinfo{pages}{543} (\bibinfo{year}{2000}).

\bibitem[{\citenamefont{Cheong et~al.}(2002)\citenamefont{Cheong, Fujisawa,
  Hayashi, Hirayama, and Jeong}}]{CheetalAPL02}
\bibinfo{author}{\bibfnamefont{H.~D.} \bibnamefont{Cheong}},
  \bibinfo{author}{\bibfnamefont{T.}~\bibnamefont{Fujisawa}},
  \bibinfo{author}{\bibfnamefont{T.}~\bibnamefont{Hayashi}},
  \bibinfo{author}{\bibfnamefont{Y.}~\bibnamefont{Hirayama}}, \bibnamefont{and}
  \bibinfo{author}{\bibfnamefont{Y.~H.} \bibnamefont{Jeong}},
  \bibinfo{journal}{Appl.\ Phys.\ Lett.} \textbf{\bibinfo{volume}{81}},
  \bibinfo{pages}{3257} (\bibinfo{year}{2002}).

\end{thebibliography}

\end{document}